\documentclass[journal=jacsat,manuscript=article]{achemso}

\usepackage{amsmath,amssymb}
\usepackage{graphicx}
\usepackage{float}
\usepackage{siunitx}
\usepackage{hyperref}
\usepackage{caption}
\usepackage{subcaption}
\usepackage{booktabs}
\usepackage{multirow}
\usepackage{xcolor}
\usepackage{microtype}
\usepackage{amsmath, amsfonts, amssymb}
\usepackage[version=4]{mhchem}
\usepackage{etoolbox}
\usepackage[section]{placeins}

\DeclareSIUnit\angstrom{\text{\AA}}

\apptocmd{\abstract}{\sloppy}{}{}

\title{Multi-Fidelity Computational Screening of High-Entropy MBenes for
  Electrochemical \ce{CO2} Reduction}

\author{Sree Harsha Bharadwaj H}
\affiliation{Department of Materials Engineering,
  Indian Institute of Technology Gandhinagar, Gujarat 382355, India}

\author{Raghavan Ranganathan}
\affiliation{Department of Materials Engineering, Indian Institute of Technology Gandhinagar, Gujarat 382355, India}

\email{rraghav@iitgn.ac.in}

\begin{document}

\begin{abstract}
High-entropy MBenes (HE-MBenes) represent a promising, unexplored class of 2D materials for electrocatalysis. In this work, we present a systematic computational screening of 56 equiatomic quinary HE-MBene compositions from the {Ti, V, Cr, Mo, Nb, Ta, Zr, Hf} pool for CO$_2$ adsorption and electroreduction. Using the Monte Carlo Special Quasirandom Structure (MCSQS) algorithm, we generated disordered M$_1B_1$-type supercells and assessed structural stability via DFT (PBE+D3) in VASP. Of the 56 candidates, 55 passed relaxation, with 45 exhibiting negative formation energies, confirming thermodynamic stability.

To efficiently screen CO$_2$ adsorption across disordered surfaces, we developed a machine-learning interatomic potential (MLIP) using the MACE architecture. Fine-tuned on our DFT dataset, the model achieved energy RMSEs of 3.49 and 3.0 meV/atom for adsorbed and pristine sets, respectively. Active sites were identified via PDOS analysis, matching metal d-orbital signatures with CO$_2$ molecular orbitals. The rate-determining step of the CO$_2$-to-CO pathway was evaluated using the computational hydrogen electrode (CHE) model. Short-time structural integrity was assessed via AIMD at 500 K over 2.5 ps; phonon-based stability remains a priority for future work. Our results establish an integrated DFT-MLIP-AIMD framework for the rational design of high-entropy 2D materials tailored for CO$_2$ conversion.

\end{abstract}

\section{Introduction}

The unabated combustion of fossil fuels has driven atmospheric \ce{CO2} concentrations to approximately 429~ppm by the beginning of 2026~\cite{noaa_co2_trends}, imposing increasing pressure on the scientific community to develop scalable carbon capture and use strategies. Electrochemical reduction of \ce{CO2} (\ce{CO2}RR) occupies a central position in this effort because it can be powered by renewable electricity, operates under mild aqueous conditions, and offers the prospect of tunable product selectivity through the choice of applied potential and catalyst composition~\cite{di2025,huynh2026}. Depending on the number of steps of proton-electron transfer steps (2 to 12), \ce{CO2}RR produces a range of value-added C$_1$ products, including carbon monoxide (CO), formic acid (HCOOH), methanol (\ce{CH3OH}), and methane (\ce{CH4}), as well as long-chain \ce{C2}--\ce{C3} serving as renewable energy carriers or industrial feedstocks~\cite{di2025}.

Despite decades of research, the practical implementation of \ce{CO2}RR remains severely constrained by the intrinsic limitations of conventional single-metal electrocatalysts. The \ce{CO2} molecule is thermodynamically inert (C=O bond dissociation energy 532~kJ~mol$^{-1}$; overall dissociation enthalpy $\Delta H_{\rm diss} \approx 1598$~kJ~mol$^{-1}$), and its activation requires, a concerted proton-coupled multi-electron transfer process~\cite{di2025}. On transition metal surfaces, the adsorption energies of key intermediates (\ce{COOH^*}, \ce{CO^*}, \ce{CHO^*}) are governed by linear scaling relationships (LSRs) that tie all binding strengths to a single descriptor, typically the CO adsorption energy~\cite{bai2023,chen2022}. This thermodynamic coupling imposes a fundamental volcano-top ceiling: any perturbation that strengthens binding of one intermediate simultaneously weakens others, precluding independent optimization of activity and selectivity. Noble metals (Au, Ag) produce CO selectively, but only at high overpotentials ($-0.9$ to $-1.2$~V vs.\ RHE), while copper, the only elemental metal capable of reducing \ce{CO2} to hydrocarbons, operates at even more negative potentials ($<-1.0$~V vs.\ RHE) with poor product selectivity~\cite{bai2023}. 
These limitations have driven extensive research into catalyst architectures capable of avoiding LSRs while simultaneously achieving low overpotential, high selectivity, and long-term stability.

Two-dimensional (2D) materials represent a compelling alternative platform for \ce{CO2}RR, combining exceptionally high surface-to-volume ratios, tunable electronic structures, and surface chemistries distinct from their bulk counterparts~\cite{huynh2026,xiao2020nano}. Among 2D materials, MXenes, the family of transition metal carbides, nitrides, and carbonitrides, have been extensively investigated due to their high electrical conductivity (6000--8000~S~cm$^{-1}$) and large accessible surface area. However,a critical limitation is the propensity of MXenes to form stable terminal oxygen and hydroxyl groups under aqueous operating conditions, progressively passivating active sites and degrading \ce{CO2}RR performance~\cite{lund2025}. This limitation has prompted growing interest in the boron analogs of MXenes, namely two-dimensional transition metal borides or MBenes~\cite{zheng2025,lu2023}. In MBenes, the boron sublattice provides a reservoir of surface lone-pair electrons that donate charge to the surface transition metal atoms, increasing the local electron density and facilitating back-donation into the \ce{CO2} $\pi^*$ orbital, thereby activating the otherwise inert molecule~\cite{li2024,lu2023}. Initially theoretically predicted  and later realized by selective etching of layered ternary MAB phases, MBenes are characterized by high electrical conductivity, notable mechanical stiffness, appreciable chemical stability, and crucially, a surface chemistry not dominated by terminal oxygen groups~\cite{zheng2025,yuan2019diboride,gunda2021progress,khaledi2021hexagonal}.

The broad electrocatalytic versatility of this emerging materials class is further underscored by recent experimental advances in their top-down exfoliation. For instance, the successful synthesis of vacancy-rich \ce{TiB2} nanosheets has unlocked highly active interfaces capable of driving the electrochemical nitrogen reduction reaction (NRR) with exceptional ammonia yields~\cite{rasyotra2024tib2}.

Early computational investigations established MBenes as highly promising \ce{CO2}RR catalysts. Lu \textit{et al.} demonstrated that MoB MBene activates \ce{CO2} with an interaction energy of $-3.64$~eV at the bridge site, exceeding the \ce{Mo2C} MXene benchmark, while effectively suppressing the competing hydrogen evolution reaction (HER)~\cite{lu2023}. Xiao \textit{et al.} identified specific \ce{TM_xB_y} compositions with high selectivity for C$_1$~\cite{xiao2021}, and Li \textit{et al.} revealed that \ce{M3B4}-type MBenes achieve lower limiting potentials than MXenes by exploiting the lone-pair surface electron activation mechanism~\cite{li2024}. Peng \textit{et al.} reported that \ce{Cr2B3} exhibits a limiting potential of $-0.31$~V for the conversion of \ce{CO2} to \ce{CH4}~\cite{peng2025}, and Di \textit{et al.} demonstrated that hexagonal MBenes terminated with hydroxyl can achieve limiting potentials as low as $-0.46$~V (ScBOH) towards the \ce{CO2}RR~\cite{di2025}. Most recently, Bai \textit{et al.} showed that single-atom embeddings in \ce{Mo2B2} break intermediate scaling relationships, producing limiting potentials of $-0.32$~V (\ce{CH4}) and $-0.27$~V (\ce{CH3OH})~\cite{bai2023}. Concurrently on the experimental front, Rasyotra \textit{et al.} successfully synthesized \ce{TaB2} nanosheets via surfactant-assisted exfoliation. Crucially, these quasi-2D metal diborides catalyzed the reduction of \ce{CO2} to ethylene (\ce{C2H4}) with an exceptional Faradaic efficiency of 75\% at $-0.85$~V vs.\ RHE, demonstrating the unique capability of \ce{Ta-B} active sites to facilitate C--C coupling~\cite{rasyotra2025_co2}. Although these results are encouraging, all the best-performing conventional MBene catalysts either target multi-electron products (CH$_4$, CH$_3$OH) at finite overpotential or achieve low overpotentials only for specific functional group terminations, motivating the search for architectures capable of driving the technologically important two-electron \ce{CO2}-to-CO pathway spontaneously.

In parallel, high-entropy alloys (HEAs), multicomponent systems comprising five or more principal elements in nearly equimolar ratios, have emerged as a paradigm-changing catalyst class~\cite{zhu2025sm,zhu2025hea_afm}. The four core effects of HEAs (thermodynamic stabilization by high configurational entropy, severe lattice distortion, sluggish atomic diffusion, and the cocktail effect) create a heterogeneous surface landscape whose wide distribution of local chemical environments collectively decouples intermediate adsorption energies, circumvents LSR constraints, and enables simultaneous optimization of multiple elementary step energetics~\cite{pedersen2020,chen2022}. 
Pedersen \textit{et al.} first demonstrated computationally that HEA compositions can be tuned to favor CO-strong, H-weak sites that suppress HER while enhancing \ce{CO2} reduction~\cite{pedersen2020}. Machine-learning-accelerated DFT studies have since screened thousands of configurations and identified selective HEA candidates, including FeCoNiCuMo with limiting potentials in the range $-0.29$ to $-0.51$~V, for multi-electron \ce{CO2}RR products~\cite{chen2022,rittiruam2024,rittiruam2025,wu2022,wang2021,yu2025}.

\begin{figure}[htbp]
    \centering
    \includegraphics[height=0.35\textheight, keepaspectratio]{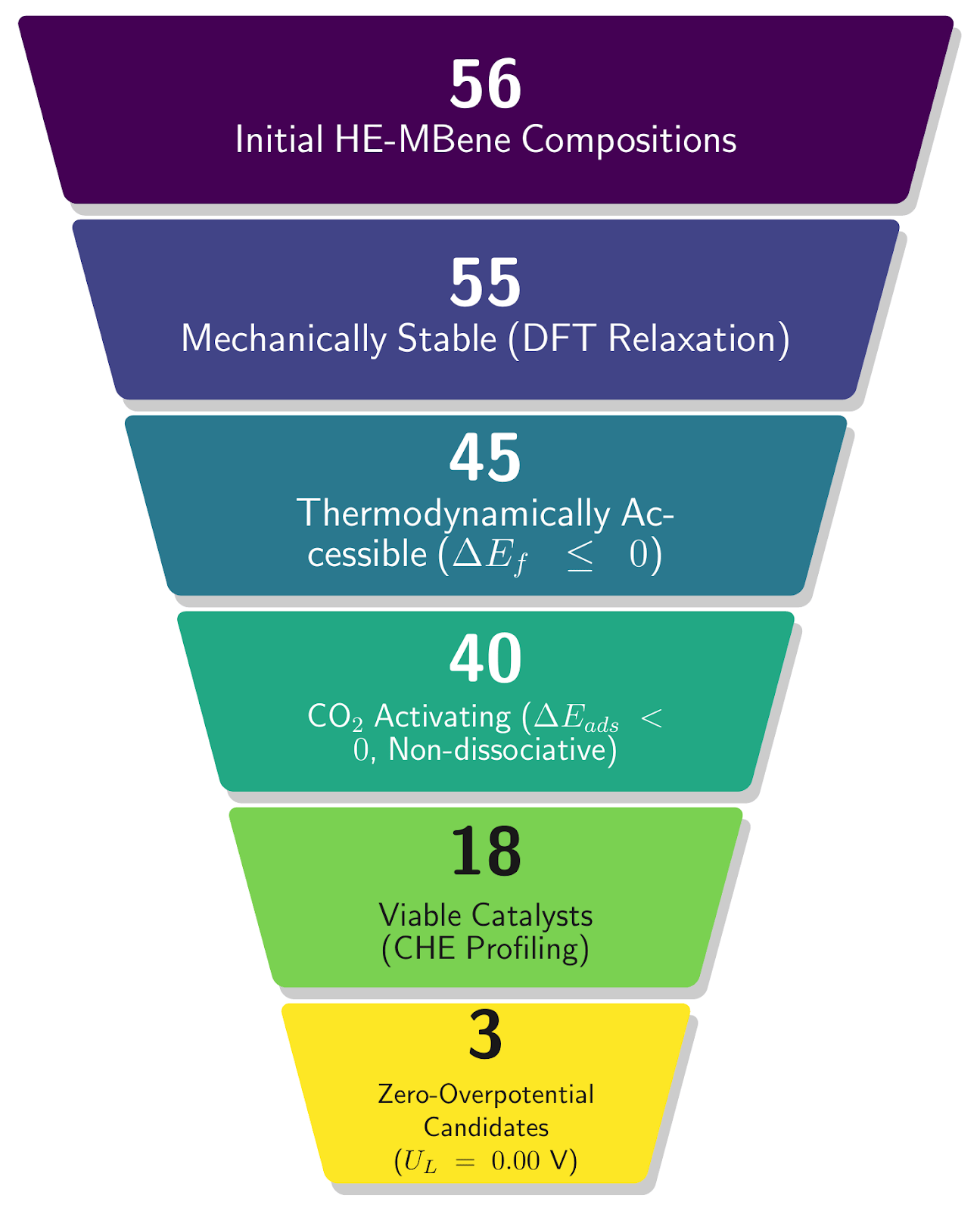}
    \caption{Multi-fidelity computational screening workflow demonstrating the progressive filtering of 56 initial equiatomic quinary HE-MBene compositions down to 3 zero-overpotential candidates.}
    \label{fig:screening_funnel}
\end{figure}

Despite rapid and largely independent progress in MBene catalysis and high-entropy materials, the intersection of these two fields, namely the use of high-entropy MBenes as two-dimensional electrocatalysts, has not been systematically explored. The equiatomic mixing of multiple transition metals on the MBene metal sublattice constitutes a natural and strategically motivated union of two powerful design principles: the surface lone-pair-driven \ce{CO2} activation intrinsic to MBenes, and the compositional diversity of the cocktail effect that decouples intermediate adsorption energies and breaks LSRs. We address this gap through a multi-fidelity computational screening of all 56 equiatomic quinary HE-MBene compositions accessible from the \{Ti, V, Cr, Mo, Nb, Ta, Zr, Hf\} elemental pool. The workflow integrates MCSQS structural generation, multi-stage DFT stability and electronic characterization, PDOS-guided adsorption site identification, MACE-MLIP-accelerated adsorption screening of 1,375 configurations, CHE free energy profiling, and AIMD thermal validation, successfully narrowing 56 candidates to 18 viable compositions, three of which achieve zero-overpotential \ce{CO2}-to-CO conversion, as outlined in Fig.~\ref{fig:screening_funnel}. Beyond identifying specific catalytic candidates, our analysis extracts universal compositional design rules, namely the electron-donor roles of Hf/Zr and the Lewis-acid active site function of Cr, which generalize to the broader high-entropy 2D materials space.
\section{Computational Details}

\subsection{Structure Generation}

High-entropy MBene supercells were constructed in the  M$_1$B$_1$-type stoichiometric lattice ($a = 3.3557$~\AA{}, $c = 4.7820$~\AA{}; space group P6/mmm), which places two symmetrically equivalent metal layers at Wyckoff positions $z = 0.748$ and $z = 0.252$, sandwiching a boron sublattice at $z = 0.500$. The chemical disorder was introduced using the Monte Carlo Special Quasirandom Structure (MCSQS) algorithm implemented in the Alloy Theoretic Automated Toolkit (ATAT)~\cite{vandewalle2002,vandewalle2013}. A $5 \times 5 \times 1$ supercell (25 metal sites per layer; 100 atoms total: 50 metal and 50 boron) was adopted with equal occupancy probabilities of 0.20 across all five constituent metal elements. Cluster correlations were evaluated using \texttt{corrdump} with a cutoff radius $R^{\rm cut} = 3.35$~\AA{}, which captures two crystallographically distinct nearest-neighbor metal-metal pair shells at 2.376~\AA{} (intra-layer) and 2.406~\AA{} (inter-layer). The terms of three-body cluster were excluded as a crystallographic necessity: the intervening boron sublattice displaces all metal-metal-metal distances beyond $R^{\rm cut}$. Pair correlations are the leading-order descriptors of chemical short-range order in the cluster expansion formalism~\cite{sanchez1984}, which directly govern the Warren-Cowley parameters~\cite{cowley1950,cowley1965} that control mixing enthalpy, phase stability, and configurational entropy. Monte Carlo optimization was run for 100,000 steps with a fixed random seed; MCSQS reproduced all 20 pair correlations in both nearest neighbor shells ($\Delta\rho = 0$) for all 56 compositions, confirming that the supercell faithfully represents a perfectly random equiatomic alloy at the pair-interaction level.

\subsection{Density Functional Theory Calculations}

All Density Functional Theory (DFT) calculations were performed using the Vienna Ab initio Simulation Package (VASP)~\cite{vasp1,vasp2} within the projector augmented wave (PAW) framework~\cite{paw}. The Perdew-Burke-Ernzerhof (PBE) generalized gradient approximation was employed for the exchange-correlation functional~\cite{pbe}. To accurately account for weak adsorption interactions, dispersion corrections were included using the DFT-D3 scheme with Becke-Johnson damping~\cite{grimme}. A plane-wave kinetic energy cutoff of 550~eV was used in conjunction with a $\Gamma$-centered $2 \times 2 \times 1$ Monkhorst-Pack $k$-point mesh (corresponding to a $k$-spacing of 0.040~\AA$^{-1}$). Electronic occupations were smeared using the first-order Methfessel-Paxton method ($\sigma = 0.2$~eV).

To capture the magnetic ground states reliably, all calculations were fully spin-polarized with initial magnetic moments of $3.0~\mu_{\text{B}}$ assigned to all metal sites.

To accurately capture the magnetic ground states, all calculations were fully spin-polarized with initial magnetic moments of $3.0~\mu_{\text{B}}$ assigned to all metal sites. A vacuum layer of 20~\AA{} was introduced along the $z$-direction to prevent spurious interactions between periodic images. The work function ($\Phi$) was determined as:
\begin{equation}
    \Phi = E_{\text{vac}} - E_F,
\end{equation}
where $E_{\text{vac}}$ represents the electrostatic potential at the vacuum plateau and $E_F$ is the Fermi level, both extracted from the planar-averaged electrostatic potential profile (see Figure~\ref{fig:pdos_wf}, right panels). 

Structural relaxations followed a rigorous three-stage coarse-to-fine protocol—incorporating ionic relaxation, cell geometry optimization, and combined ionic/cell-shape refinement—to ensure convergence to global minima. Electronic steps and ionic relaxations were converged to thresholds of the order of $10^{-3}$~eV and $10^{-2}$~eV/\AA, respectively. Out of the 56 initial candidates, 55 structures reached mechanically stable configurations; HES~ID-9 was found to be unstable and was excluded.

The thermodynamic stability of the remaining 55 structures was assessed via the formation energy per atom ($\Delta E_f$):
\begin{equation}
  \Delta E_f = \frac{E_{\text{HE-MBene}} - \sum_{i} n_i E_i}{\sum_{i} n_i},
  \label{eq:formation}
\end{equation}
where $E_{\text{HE-MBene}}$ is the total energy of the supercell, $E_i$ is the reference energy of species $i$ in its stable bulk phase, and $n_i$ is the number of atoms of species $i$. Compositions exhibiting $\Delta E_f > 0$ were deemed thermodynamically inaccessible, reducing the candidate pool to 45 structures (excluded IDs: 1, 2, 5, 11, 21, 36, 37, 39, 42, 46, and 51; see Table~\ref{tab:pristine-compositions}).

For the 45 stable structures, the projected density of states (PDOS) was analyzed. The preferred adsorption site was identified by locating the surface metal element exhibiting the highest peak $d$-orbital intensity (electronic resonance) within the energy window of $-2$ to $+1$~eV relative to $E_F$. This ``Peak-in-Window" criterion was selected to target the maximum density of local states specifically  aligned with the $\pi$ and $\pi^*$ frontier molecular orbitals of \ce{CO2}. While a total overlap integral measures overall hybridization capacity, the peak intensity identifies the specific resonance channels most efficient for back-donation into the \ce{CO2} LUMO, which is a critical driver for the initial activation of the inert molecule. Subsequent \ce{CO2} adsorption screening was performed at a validated reduced precision (500~eV cutoff). This protocol was cross-validated against high-precision calculations for HES~ID-54 and HES~ID-29, showing a systematic shift of $< 0.7$~eV that does not affect relative stability trends.

The \ce{CO2} adsorption energy ($\Delta E_{\text{ads}}$) was calculated as:
\begin{equation}
    \Delta E_{\text{ads}} = E_{\text{slab+CO}_2} - E_{\text{slab}} - E_{\text{CO}_2}.
\end{equation}

Both on-top and hollow (bridge) sites were evaluated, with on-top configurations proving most exothermic for the majority of structures (Table~\ref{tab:co2_ads_bridge}). Following relaxation, the configurations were classified into two groups: (i) activated molecular adsorption (bent \ce{CO2^{\delta-}} with O--C--O angles of 120--150$^\circ$) and (ii) spontaneous dissociation (CO + O). Dissociative and non-binding ($\Delta E_{\text{ads}} \geq 0$) structures were excluded, resulting in a final dataset of 18 compositions for subsequent Computational Hydrogen Electrode (CHE) analysis.

\subsection{Computational Hydrogen Electrode and Selectivity Analysis}

The selectivity towards \ce{CO2}RR over the competing hydrogen evolution 
reaction (HER) was assessed by computing the Gibbs free energy of hydrogen 
adsorption ($\Delta G_{\text{H}^*}$) via the Volmer step:
\begin{equation}
    {*} + \text{H}^+ + e^- \rightarrow \text{H}^*, \quad 
    \Delta G_{\text{H}^*} = \Delta E_{\text{H}^*} + \Delta\text{ZPE} - T\Delta S.
\end{equation}
A surface is operationally selective for \ce{CO2}RR when 
$\Delta G_{\text{COOH}^*} < \Delta G_{\text{H}^*}$, indicating that the first 
protonation step of \ce{CO2} is thermodynamically preferred over competitive 
hydrogen adsorption at the active site.

The two-electron \ce{CO2}-to-CO free energy pathway was evaluated using the 
computational hydrogen electrode (CHE) model of N{\o}rskov \textit{et al.}~\cite{norskov2004}:
\begin{align}
    \text{CO}_2^* + \text{H}^+ + e^- &\rightarrow \text{COOH}^*, \\
    \text{COOH}^* + \text{H}^+ + e^- &\rightarrow \text{CO}^* + \text{H}_2\text{O}.
    \label{eq:step2}
\end{align}

Within the CHE framework, the electrochemical potential of a proton-electron pair is set to $\mu(\text{H}^+ + e^-) = \tfrac{1}{2}\mu(\text{H}_2)$ at $\text{pH} = 0$ and $U = 0$~V vs.\ RHE, allowing the free energy of each elementary step to be obtained directly from DFT energies and thermodynamic corrections. Zero-point energy (ZPE) and thermal entropy ($T\Delta S$) corrections were applied at 298.15~K and 1~bar to all raw DFT energies, with reference molecular energies of $E(\text{CO}_2) = -22.96$~eV and $E(\text{H}_2) = -6.69$~eV.

The second elementary step (Eq.~\ref{eq:step2}), corresponding to the 
hydrogenation of \ce{COOH^*} to produce \ce{CO^*} and \ce{H2O}, was identified as the rate-determining step (RDS) for all compositions in this study, consistent with the established behaviour of transition metal surfaces in the two-electron \ce{CO2}RR pathway~\cite{di2025}. The thermodynamic limiting potential is accordingly defined as $U_L = -\Delta G_{\text{RDS}} / e$. 
Compositions whose RDS exhibited an upward free energy step under zero applied bias were excluded at this final screening gate, reducing the candidate set from 40 to 18 viable compositions.

\subsection{MACE Machine-Learning Interatomic Potential}

Exhaustive DFT-level \ce{CO2} adsorption sampling across all surface sites and all 40 post-screening compositions would be computationally prohibitive. 
A machine-learning interatomic potential (MLIP) based on the MACE 
architecture~\cite{mace} was therefore developed to enable high-throughput 
adsorption screening. The \texttt{mace-mh-1} foundation model, pre-trained on the Materials Project~\cite{jain2013materialsproject}and OpenMaterials~\cite{barrosoluque2024omat24} datasets, was fine-tuned on our DFT \ce{CO2}/HE-MBene dataset using the \texttt{omat\_pbe} foundation head~\cite{mace}. The fine-tuning dataset comprised 4,949 training and 620 validation configurations, spanning pristine slabs and on-top site \ce{CO2} adsorption energies and forces.

The \texttt{ScaleShiftMACE} variant was trained with RMS-force scaling for up to 200 epochs (batch size 1; learning rate $3 \times 10^{-4}$;  \texttt{ReduceLROnPlateau} scheduler with patience 10; stochastic weight  averaging from epoch 150). Energy and force loss weights were set to 1.0 and 100.0, respectively. The model was trained in double precision on a  CUDA-enabled GPU for species B, C, O, Ti, V, Cr, Zr, Nb, Mo, Ta, and Hf. The trained potential achieved chemical accuracy with an RMSE of  3.49~meV/atom, RMSE $\approx 38$~meV\,\AA$^{-1}$ and $R^2 = 0.9991$ on the held-out validation set, enabling  reliable adsorption screening across 1,375 unique surface environments.

The dynamical stability of all 54 mechanically stable HE-MBene structures was assessed by computing phonon dispersions using the finite displacement method as implemented in the \texttt{phonopy} package~\cite{phonopy}. Force constants were evaluated using the fine-tuned MACE potential rather than DFT, providing the force accuracy required for reliable phonon calculations at a fraction of the computational cost.
 
Prior to phonon evaluation, each structure was fully relaxed using the MACE calculator in a two-stage protocol: a preliminary BFGS relaxation ($f_{\rm max} = 0.05$~eV/\AA{}, 1000 steps) followed by a fine LBFGS relaxation with simultaneous cell optimization via the \texttt{FrechetCellFilter}, converged to $f_{\rm max} = 10^{-3}$~eV/\AA{} (up to 2000 steps).
All relaxations and force evaluations were performed in double precision (\texttt{float64}) to match the precision of the MACE training protocol. 
Phonon supercells were constructed using a $3 \times 3 \times 1$ expansion of the relaxed unit cell with a finite displacement amplitude of 0.01~\AA{}.
Forces were evaluated on all symmetry-inequivalent displaced supercells, with translational drift removed from each force set prior to assembly.
Force constants were symmetrized iteratively for three passes using
\texttt{phonopy}'s built-in symmetrization routine to enforce translational
and point-group invariance.

Phonon band structures were computed along the high-symmetry path
$\Gamma \to \mathrm{M} \to \mathrm{K} \to \Gamma$ with 101 points per
segment, and the phonon density of states was evaluated on a
$21 \times 21 \times 1$ $q$-point mesh. A structure was classified as dynamically stable if all phonon frequencies across the full Brillouin zone remained non-negative, with a numerical tolerance of $-0.20$~THz to exclude spurious acoustic branch artifacts arising from finite supercell size. The workflow was parallelized across multiple CUDA-enabled GPUs using a rank-based distribution scheme, with each worker processing an independent subset of structures.

\subsection{Ab Initio Molecular Dynamics (AIMD)}

To assess the thermal stability of the selected high-entropy MBene structures (IDs: 13, 25, 30, 36, and 54), ab initio molecular dynamics (AIMD) simulations were performed. The simulations were conducted at 500~K in the $NVT$ ensemble using the Nosé--Hoover thermostat. A time step of 0.2~fs was employed, and each trajectory was propagated for a total duration of 2.5~ps, corresponding to 12,500 steps.

The calculations were carried out using the VASP package with the same PAW-PBE+D3 functional settings as the static DFT calculations. To improve computational efficiency while maintaining adequate accuracy for stability assessment, a reduced $\Gamma$-centered $1 \times 1 \times 1$ $k$-point mesh was utilized. The structural integrity and thermal stability were evaluated by monitoring the evolution of the total potential energy and the Root Mean Square Deviation (RMSD) as a function of simulation time. Consistent energy profiles and low RMSD values throughout the trajectories were used to confirm that the frameworks remained intact without significant structural reconstruction or dissociation.

\section{Results and Discussion}

\subsection{Compositional Space and Structural Stability}

The complete set of 56 equiatomic quinary HE-MBene compositions screened in this work is listed in Table~\ref{tab:pristine-compositions}. All compositions share the equiatomic M$_1$B$_1$ stoichiometry with 10~at.\% of each metal element in the 50-atom metal sublattice. The three candidates with zero-overpotential that survived all four screening gates are indicated in bold.
\FloatBarrier
\begin{table}[H]
  \centering
  \caption{All 56 equiatomic quinary HE-MBene compositions screened in this
    work, indexed by HES~ID. HES~ID-9 was discarded following structural
    relaxation due to mechanical instability. \textbf{Bold} HES~IDs identify the three zero-overpotential candidates that survived all four screening gates and exhibit spontaneous \ce{CO2}-to-CO conversion at zero applied potential.}
  \label{tab:pristine-compositions}
  \setlength{\tabcolsep}{4pt}
  \begin{tabular}{|c|l|c|l|}
    \hline
    \textbf{HES ID} & \textbf{Composition} &
    \textbf{HES ID} & \textbf{Composition} \\
    \hline
    1  & VMoNbCrTiB$_5$ & 29 & CrZrHfNbTiB$_5$ \\ \hline
    2  & VMoTaCrTiB$_5$ & 30 & CrZrHfTaTiB$_5$ \\ \hline
    3  & VMoZrCrTiB$_5$ & 31 & MoTaZrNbTiB$_5$ \\ \hline
    4  & VMoHfCrTiB$_5$ & 32 & MoTaHfNbTiB$_5$ \\ \hline
    5  & VNbTaCrTiB$_5$ & 33 & MoZrHfNbTiB$_5$ \\ \hline
    6  & VNbZrCrTiB$_5$ & 34 & MoZrHfTaTiB$_5$ \\ \hline
    7  & VNbHfCrTiB$_5$ & 35 & NbZrHfTaTiB$_5$ \\ \hline
    8  & VTaZrCrTiB$_5$ & 36 & CrNbTaMoVB$_5$ \\ \hline
    10 & VZrHfCrTiB$_5$ & 37 & CrNbZrMoVB$_5$ \\ \hline
    11 & VNbTaMoTiB$_5$ & 38 & CrNbHfMoVB$_5$ \\ \hline
    12 & VNbZrMoTiB$_5$ & 39 & CrTaZrMoVB$_5$ \\ \hline
    13 & VNbHfMoTiB$_5$ & 40 & CrTaHfMoVB$_5$ \\ \hline
    14 & VTaZrMoTiB$_5$ & 41 & CrZrHfMoVB$_5$ \\ \hline
    15 & VTaHfMoTiB$_5$ & 42 & CrTaZrNbVB$_5$ \\ \hline
    16 & VZrHfMoTiB$_5$ & 43 & CrTaHfNbVB$_5$ \\ \hline
    17 & VTaZrNbTiB$_5$ & 44 & CrZrHfNbVB$_5$ \\ \hline
    18 & VTaHfNbTiB$_5$ & 45 & CrZrHfTaVB$_5$ \\ \hline
    19 & VZrHfNbTiB$_5$ & 46 & MoTaZrNbVB$_5$ \\ \hline
    20 & VZrHfTaTiB$_5$ & 47 & MoTaHfNbVB$_5$ \\ \hline
    21 & CrNbTaMoTiB$_5$ & 48 & MoZrHfNbVB$_5$ \\ \hline
    \textbf{22} & \textbf{CrNbZrMoTiB$_5$} & 49 & MoZrHfTaVB$_5$ \\ \hline
    23 & CrNbHfMoTiB$_5$ & 50 & NbZrHfTaVB$_5$ \\ \hline
    24 & CrTaZrMoTiB$_5$ & 51 & MoTaZrNbCrB$_5$ \\ \hline
    25 & CrTaHfMoTiB$_5$ & 52 & MoTaHfNbCrB$_5$ \\ \hline
    26 & CrZrHfMoTiB$_5$ & \textbf{53} & \textbf{MoZrHfNbCrB$_5$} \\ \hline
    27 & CrTaZrNbTiB$_5$ & \textbf{54} & \textbf{MoZrHfTaCrB$_5$} \\ \hline
    28 & CrTaHfNbTiB$_5$ & 55 & NbZrHfTaCrB$_5$ \\ \hline
     &  & 56 & NbZrHfTaMoB$_5$ \\ \hline
  \end{tabular}
\end{table}
\FloatBarrier
\subsection{Electronic Structure and Thermodynamic Properties}

\begin{figure}[H]
  \centering
  \includegraphics[width=0.78\linewidth]{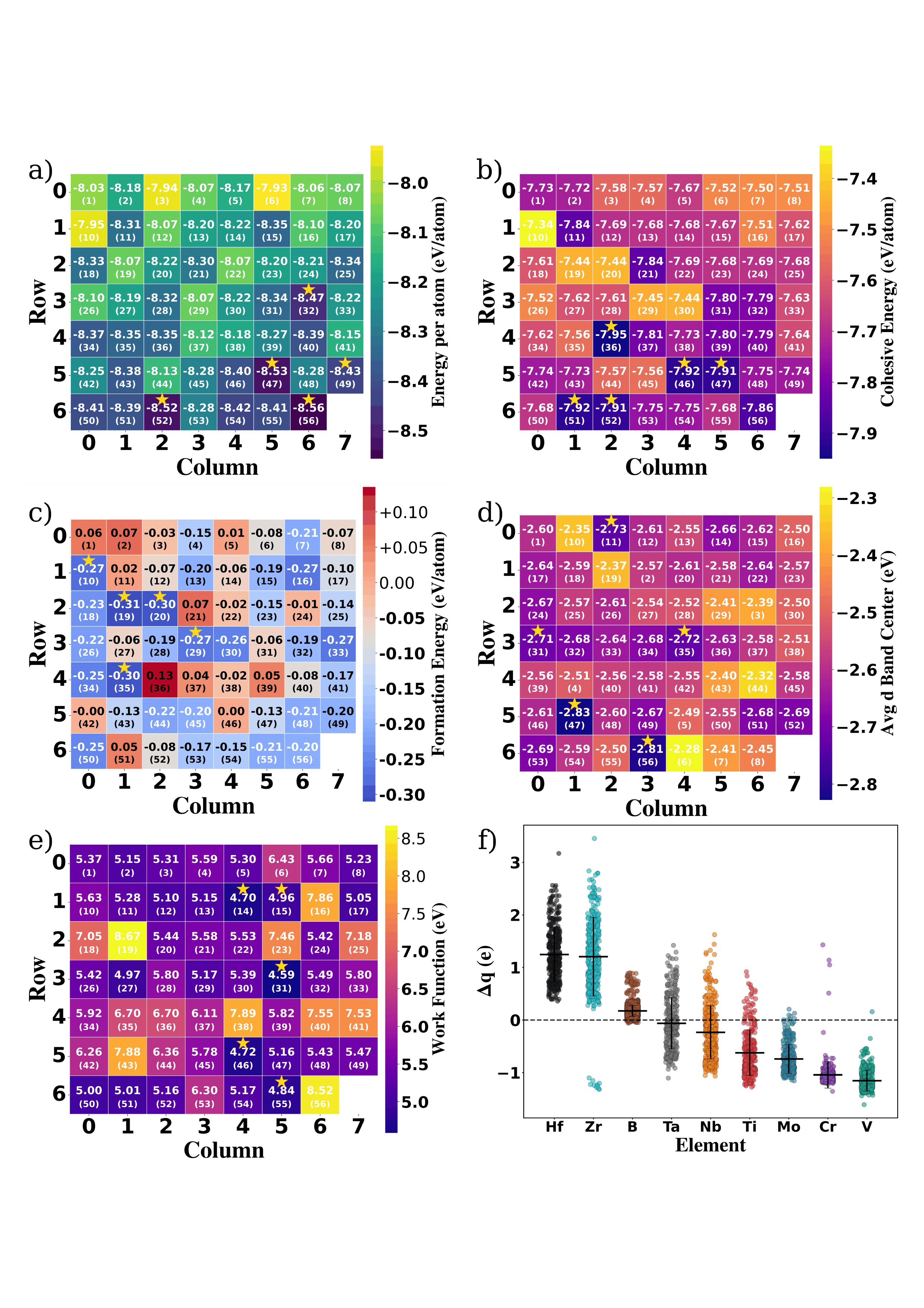}
  \caption{\textbf{Compositional property maps for all 56 equiatomic quinary
    HE-MBene compositions computed from DFT.}
    (a) Ground-state energy per atom (eV/atom). (b) Cohesive energy (eV/atom). (c) Formation energy per atom (eV/atom); compositions above     zero (red cells) are thermodynamically unstable and were excluded from further analysis. (d) Average $d$-band centre relative to $E_F$ (eV); the narrow approximately 0.55~eV spread is a fingerprint of the high-entropy cocktail effect. (e) Work function (eV); the greater than 3~eV compositionally tunable range illustrates the exceptional electronic versatility of HE-MBenes. (f) Element-resolved Bader charge ($\Delta q = N_{\rm valence} - N_{\rm Bader}$, in $e$), revealing the Hf/Zr electron-donor and V/Cr electron-acceptor roles that underpin \ce{CO2} activation. Stars ($\star$) mark global extrema within each panel; HES~IDs are given in parentheses.}
  \label{fig:drawing}
\end{figure}

Figure~\ref{fig:drawing} presents six compositional property maps derived from DFT calculations for all 56 HE-MBene compositions. Together, these maps provide a holistic electronic and thermodynamic fingerprint of the compositional space and reveal the physical origins of stability and catalytic activity. 

\textbf{Energy in the ground-state.} The energy in the ground state per atom (Figure~\ref{fig:drawing}a) ranges $-7.93$ to $-8.56$~eV/atom throughout the compositional space, with the global minimum at HES~ID-56 ($-8.56$~eV/atom) and the highest value at HES~ID-6 ($-7.93$~eV/atom). These strongly negative values are consistent with the high cohesive stability previously reported for 2D transition-metal diborides and hexagonal MBenes~\cite{yuan2019diboride,khaledi2021hexagonal}. The approximately 0.6~eV/atom spread over 56 compositions indicates that the collective multicomponent interaction, rather than any single elemental substitution, governs cohesive stability. The cohesive energy distribution (Figure~\ref{fig:drawing}b) closely mirrors the total energy map ($-7.34$ to $-7.95$~eV/atom), confirming that the high-entropy mixing on the MBene scaffold does not compromise structural integrity. This observation is consistent with the high-entropy stabilization principle: the large configurational entropy $T\Delta S_{\rm conf}$ (theoretical maximum $R\ln 5 \approx 0.134$~eV/atom at 500~K for equiatomic five-component systems) reduces the Gibbs free energy of mixing sufficiently to suppress phase segregation into competing binary or ternary boride phases. 

\textbf{The formation energy and thermodynamic accessibility.} The formation energies (Figure~\ref{fig:drawing}c) span from $+0.13$ to approximately $-0.31$~eV/atom. The majority compositions cluster between $-0.05$ and $-0.30$~eV/atom, confirming the wide thermodynamic accessibility in the compositional space. The most negative formation energies are concentrated among compositions containing large fractions of group-IV and group-V elements (Zr, Hf, Nb, Ta), consistent with the well-established strength of covalent metal-boron bonding in these transition metal borides~\cite{yuan2019diboride,gunda2021progress}, where the electropositive character of early transition metals maximizes ionic-covalent charge transfer to boron. Conversely, all eleven excluded compositions are enriched in group-V and group-VI elements (V, Cr, Mo) without sufficient group-IV counterbalancing, underscoring the critical stabilizing roles of Zr and Hf. This formation energy map therefore directly encodes the first compositional design rule: a minimum loading of group-IV elements is required to ensure thermodynamic stability. 

\textbf{The average $d$-band center.} The average $d$-band center values. (Figure~\ref{fig:drawing}d) span $-2.28$ to $-2.83$~eV relative to the Fermi level, a narrow range of only approximately 0.55~eV compared to the variation observed between individual pure transition metals. According to the Hammer-N{\o}rskov $d$-band model~\cite{hammer1995}, a $d$-band center closer to the Fermi level indicates stronger orbital hybridization with adsorbate states and thus stronger intermediate binding. The narrow spread across 56 compositions is a direct fingerprint of the cocktail effect: high-entropy mixing effectively averages the $d$-band contributions of all five constituent elements, producing controlled incremental shifts in $\varepsilon_d$ rather than the abrupt transitions seen between monometallic surfaces~\cite{khaledi2021hexagonal,di2025}. This averaging mechanism limits the risk of over-binding and provides a continuously tunable binding-energy lever through compositional variation, a key advantage of HE-MBenes over conventional single-element or binary MBene catalysts. 

\textbf{Work function.} Work function values (Figure~\ref{fig:drawing}e) vary by more than 3~eV throughout the compositional space, a remarkable range for a class of isomorphic 2D materials. The work function directly quantifies the thermodynamic cost of transferring an electron from the surface Fermi level to vacuum, and lower values correlate with a greater propensity for surface-to-adsorbate charge transfer during \ce{CO2} activation via $\pi^*$ back-donation. As confirmed by the planar-averaged electrostatic potential profiles in Figure~\ref{fig:pdos_wf}, substituting electron-acceptor elements (V, Cr) with group-IV donors (Zr, Hf) substantially reduces the work function across the compositional space. The three candidates for zero-overpotential (within the CHE thermodynamic framework) (HES~IDs 22, 53, 54) occupy intermediate work function values, consistent with the Sabatier principle: an excessively low work function drives, aggressive dissociative interaction \ce{CO2}, while an excessively high work function suppresses initial activation altogether. Furthermore, the correlation between high peak intensity in the PDOS and adsorption energy suggests that reactivity is governed by the availability of concentrated electronic states at resonance energies rather than the total integrated d-band density. This approach aligns with the concept of LDOS  engineering, where the magnitude of the local density of states at specific  energy levels serves as a primary descriptor for catalytic activity on complex multi-component surfaces.

\textbf{Bader charge analysis:} The element-resolved Bader charge analysis (Figure~\ref{fig:drawing}f) reveals a clear and chemically intuitive donor-acceptor hierarchy across the eight candidate elements. Hf and Zr emerge as the most pronounced electron donors (median $\Delta q \approx +1.2\,e$ and $+0.8\,e$, respectively), while V and Cr are the strongest acceptors (median $\Delta q \approx -1.0\,e$ and $-1.1\,e$). Boron occupies an intermediate donor role ($\Delta q \approx +0.3\,e$), consistent with its structural bridge position between the metal layers. Ta and Nb are near charge-neutral. This hierarchy follows naturally from the electronegativity and filling of the $d$-band across the series of transition metals and provides the primary atomistic rationalization of the \ce{CO2} adsorption landscape: Hf/Zr-rich environments generate elevated local electron densities that allow effective back-donation into the \ce{CO2} $\pi^*$ orbital, activating the molecule without inducing full dissociation~\cite{lu2023,li2024}.
\FloatBarrier
\FloatBarrier

\subsection{PDOS-Guided Active Site Identification}

\begin{figure}[tp!] 
\centering 
\includegraphics[width=0.75\linewidth]{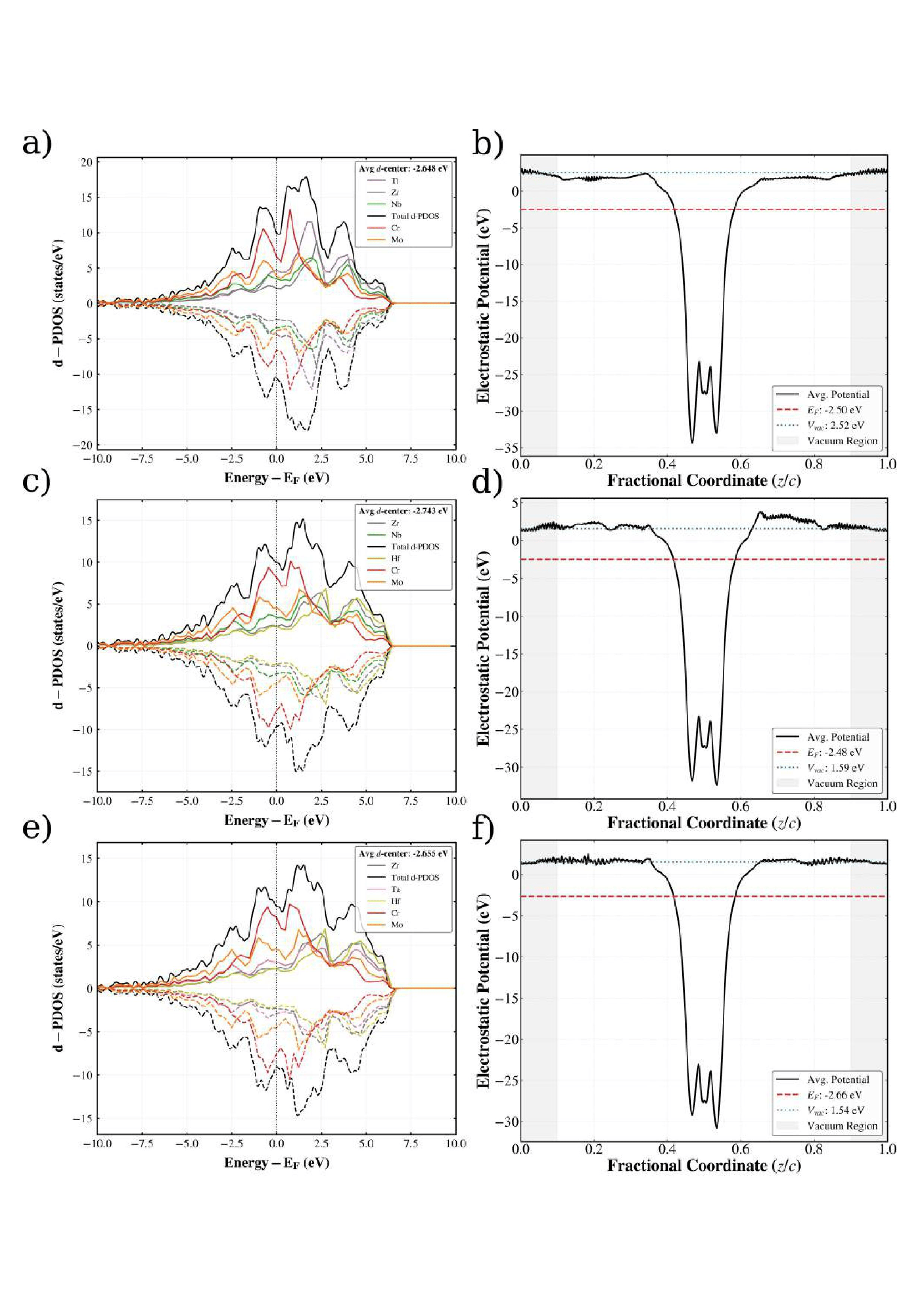} 
\caption{\textbf{Element-resolved $d$-PDOS and planar-averaged electrostatic potential for the three zero-overpotential (within the CHE thermodynamic framework) HE-MBene candidates.} Left panels (a, c, e): spin-polarised $d$-PDOS for each constituent metal element in HES~ID-22, HES~ID-53, and HES~ID-54, respectively (solid lines: spin-up; dashed lines: spin-down). The average $d$-band center $\varepsilon_d$ is annotated in each panel. The Cr $d$-manifold (red) carries the largest unoccupied spectral weight in the energy window $-2$ to $+1$~eV relative to $E_F$, the region of energetic overlap with the \ce{CO2} $\pi^*$ frontier orbital, identifying Cr as the preferred adsorption center. Right panels (b, d, f): planar-averaged electrostatic potential along the $z$-direction for the corresponding compositions. The Fermi level ($E_F$, red dashed) and vacuum level ($V_{\rm vac}$, blue dotted) are indicated; their difference yields the work function. The vacuum plateau in all three profiles confirms that the 20~\AA{} vacuum layer eliminates spurious periodic-image interactions.} 
\label{fig:pdos_wf} 
\end{figure}

Figure~\ref{fig:pdos_wf} presents the spin-polarised element-resolved
$d$-PDOS and the planar-averaged electrostatic potential for the three
zero-overpotential HE-MBene candidates (HES~IDs 22, 53, and 54). For each composition, the $d$-PDOS analysis was used to identify the metal
element whose unoccupied $d$-states contribute most strongly in the energy
window between approximately $-2$ and $+1$~eV relative to $E_F$ --- the
region that energetically overlaps with the \ce{CO2} $\pi^*$ frontier orbital.

In all three compositions, the Cr $d$-orbital manifold (red curves) exhibits substantial spectral weight just above and below $E_F$, with a prominent peak near the Fermi level and a well-developed tail in the unoccupied region above $E_F$.
In comparison, the Mo contributions (orange) are more delocalised and spread further below $E_F$, while the group-IV and group-V elements (Zr, Hf, Nb, Ta) show their dominant $d$-weight concentrated well below $-2$~eV and are thus less energetically accessible for back-donation into the \ce{CO2} $\pi^*$ orbital.
Based on this element-resolved analysis, Cr was designated as the preferred adsorption centre for all three compositions, a designation further corroborated by the adsorption heatmap data in Figure~\ref{fig:co2_landscape}, which shows Cr-labelled cells for HES~IDs 22, 53, and 54.
The average $d$-band centres ($\varepsilon_d = -2.648$, $-2.743$, and $-2.655$~eV for HES~IDs 22, 53, and 54, respectively) fall within the narrow 0.55~eV window identified from the global property map and are shifted toward the Fermi level relative to the unalloyed Cr surface
($\varepsilon_d \approx -2.9$~eV for pure Cr~\cite{hammer1995}), directly
reflecting the charge enrichment delivered by the Hf/Zr electron-donor
neighbours.

The planar-averaged electrostatic potential profiles (right panels) confirm that the 20~\AA{} vacuum layer produces a well-defined flat vacuum plateau. The work function values extracted as $\Phi = E_{\rm vac} - E_F$ from these profiles are internally consistent with the values reported in
Figure~\ref{fig:drawing}e. 

\FloatBarrier
\subsection{MACE Potential Validation and the \ce{CO2} Adsorption Landscape}

\begin{figure}[tp!] \centering \includegraphics[width=0.85\linewidth]{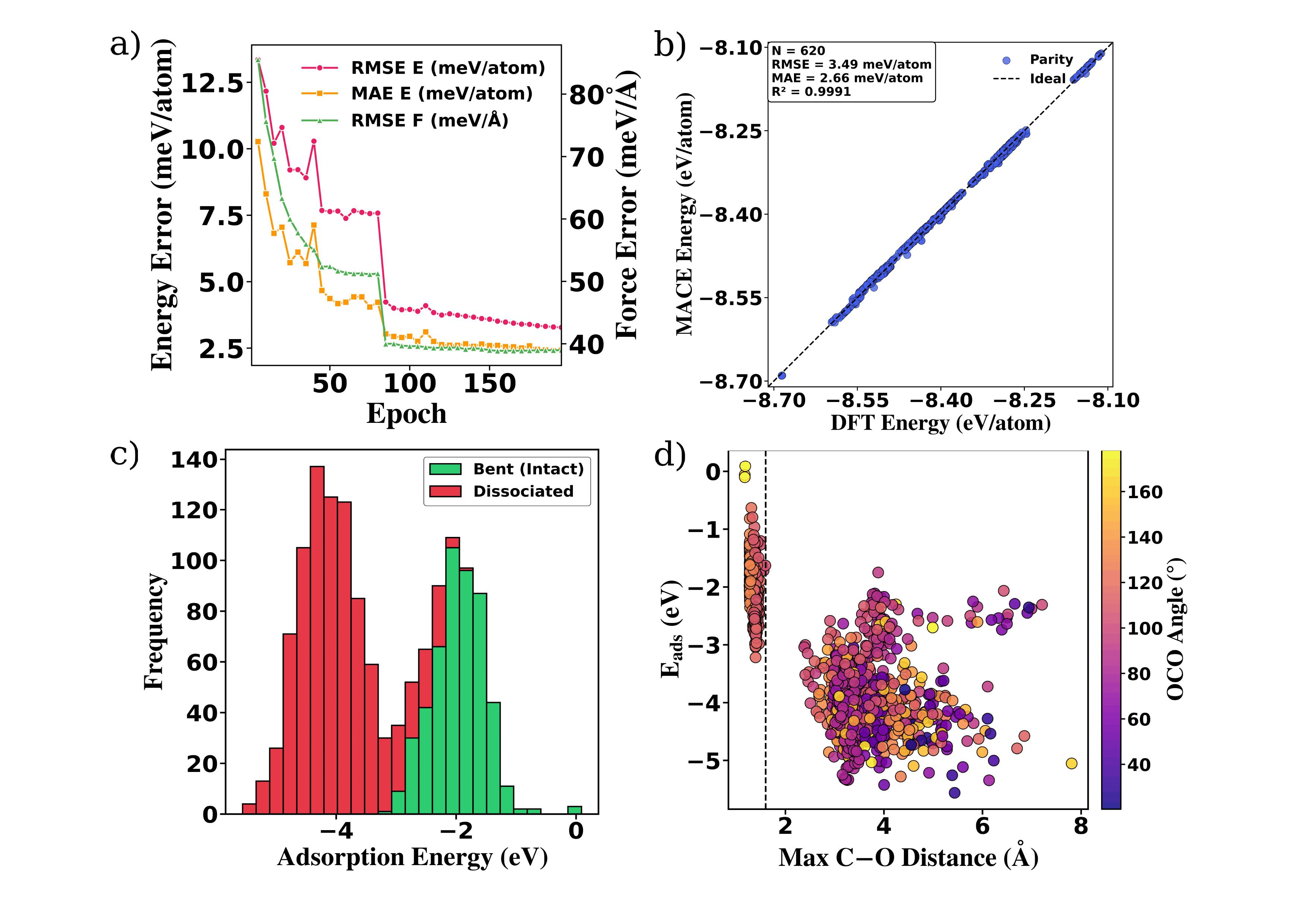} \caption{\textbf{MACE MLIP training diagnostics and global \ce{CO2} adsorption statistics.} (a) Training convergence curves showing RMSE on energies (pink circles), MAE on energies (orange squares), and RMSE on forces (green triangles) as functions of training epoch; convergence to 3.49~meV/atom energy RMSE and approximately 38~meV/\AA{} force RMSE confirms chemical accuracy. (b) Parity plot of MACE-predicted versus DFT-computed energies per atom on the held-out test set ($N = 620$); $R^2 = 0.9991$ demonstrates faithful generalisation across the full range of surface configurations. (c) Global distribution of \ce{CO2} adsorption energies ($\Delta E_{\rm ads}$) across the 1,375-configuration dataset, categorised by molecular integrity (intact bent \ce{CO2^{\delta-}} versus dissociated CO\,$+$\,O). The intact population peaks near $-2.0$~eV and the dissociated population near $-4.0$~eV; the crossover near $\Delta E_{\rm ads} \approx -2.5$~eV marks the activation-dissociation boundary. (d) Scatter plot of maximum C-O bond distance (\AA{}) versus $\Delta E_{\rm ads}$, colour-coded by O-C-O bond angle ($^\circ$). Intact \ce{CO2^{\delta-}} configurations cluster near the intact C-O distance (dashed vertical line at $\sim 1.9$~\AA{}); dissociated configurations extend to large separations. The progressive colour shift from yellow ($\sim 160^\circ$) to purple ($<60^\circ$) tracks molecular activation with increasing $|\Delta E_{\rm ads}|$ up to the dissociation threshold.} \label{fig:ml} \end{figure}

Figure~\ref{fig:ml} presents the MACE MLIP training diagnostics and the
global adsorption statistics for all 1,375 configurations evaluated during
the high-throughput screening phase.

Figure~\ref{fig:ml}a shows the evolution of energy and force prediction errors over 200 training epochs, which converge to 3.49~meV/atom (energy RMSE), 2.66~meV/atom (energy MAE), and approximately 38~meV/\AA{} (force RMSE).
The smooth, monotonic convergence without overshoot demonstrates that the \texttt{ReduceLROnPlateau} scheduler and stochastic weight averaging strategy are effective for this multi-element, multi-phase dataset.
The force RMSE is particularly relevant because the identification of the adsorption site and the relaxation of geometry depend on accurate force prediction, not merely on energy ranking.

Figure~\ref{fig:ml}b presents the parity plot on the 620-configuration held-out test set: data points cluster tightly around the $y = x$ line across the full approximately 4~eV/atom energy range with $R^2 = 0.9991$, satisfying the chemical accuracy threshold ($\leq 1$~kcal~mol$^{-1}$ $\approx 43$~meV) for MLIP benchmarking~\cite{Liu2023MLIPMetrics,Jacobs2025MLIPGuide} and confirming that the fine-tuned MACE potential generalises faithfully to configurations not represented in the training set.

The DFT benchmark against single-metal MBene references is presented in
Table~\ref{tab:co2_ads_bridge}.The close agreement between this work and published theoretical values (with a mean absolute deviation below 0.15~eV across five distinct MBene compositions) validates the exchange-correlation functional, dispersion correction, cutoff energy, and $k$-point sampling employed throughout the screening.

\begin{figure}[htbp!]
  \centering
  \includegraphics[width=0.85\linewidth]{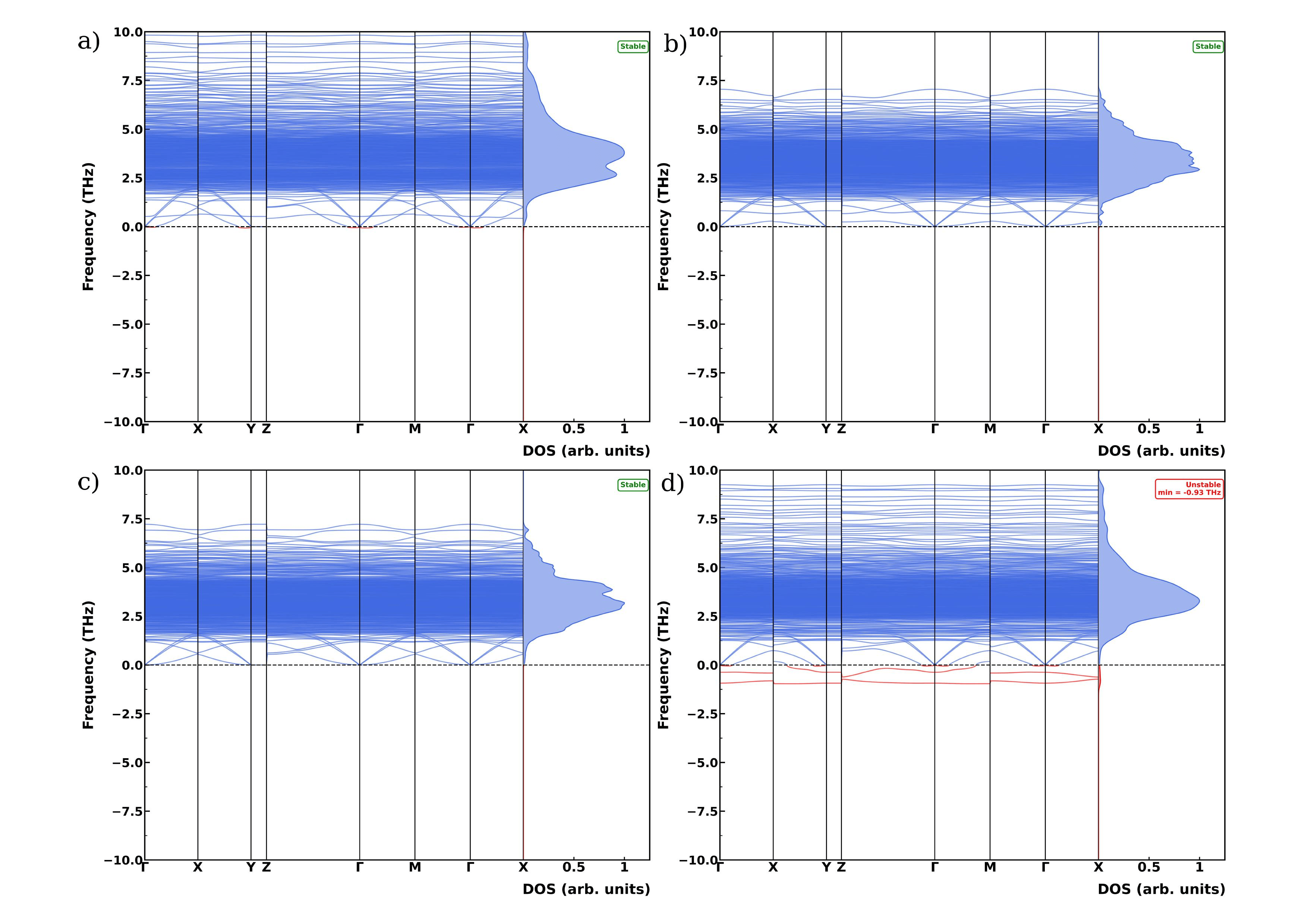}
  \caption{\textbf{Representative MACE-computed phonon band structures and
    density of states for HE-MBene compositions.}
    (a) HES~ID-22, (b) HES~ID-53, and (c) HES~ID-54 --- the three
    zero-overpotential candidates --- all showing fully positive phonon
    frequencies across the Brillouin zone, confirming dynamical stability.
    (d) HES~ID-21, a representative dynamically unstable structure
    (minimum frequency $= -0.93$~THz); imaginary modes are highlighted in
    red and dip below zero near the zone boundary, indicating a soft-mode
    instability.
    The dashed horizontal line at 0~THz marks the stability threshold.
    Band structures are plotted along the $\Gamma \to \mathrm{X} \to
    \mathrm{Y} \to \mathrm{Z} \to \Gamma \to \mathrm{M} \to \Gamma$
    high-symmetry path; the phonon DOS is shown in the right panel of
    each subfigure.}
  \label{fig:phonon}
\end{figure}

The validated MACE potential was further employed to assess the dynamical stability of all 55 mechanically stable structures via phonon dispersion calculations along the $\Gamma \to \mathrm{M} \to \mathrm{K} \to \Gamma$ high-symmetry path~\cite{phonopy}, confirming 40 dynamically stable compositions (Figure~\ref{fig:phonon}; Table~\ref{tab:phonon_unstable}). The full set of 55 structures was subsequently carried forward for \ce{CO2} adsorption screening across all 1,375 unique configurational environments, with dynamically unstable compositions excluded only at the final CHE gate.
 
\begin{table}[htbp!]
  \centering
  \caption{\textbf{Dynamically unstable HE-MBene structures identified by
    MACE-computed phonon analysis.}
    Structures are ranked by increasing severity of the imaginary mode.
    The minimum phonon frequency (THz) is taken as the most negative value
    across the $\Gamma \to \mathrm{M} \to \mathrm{K} \to \Gamma$ band path
    and the $21 \times 21 \times 1$ $q$-mesh.
    HES~IDs correspond to the composition index in
    Table~\ref{tab:pristine-compositions}.
    Structures above the dashed line passed the DFT formation energy filter
    but were excluded at this phonon gate; structures below were already
    flagged by formation energy.}
  \label{tab:phonon_unstable}
  \small
  \begin{tabular}{cc}
    \hline
    HES~ID & Minimum frequency (THz) \\
    \hline
    52     & $-0.89$ \\
    21     & $-0.93$ \\
    32     & $-0.99$ \\
    36     & $-1.02$ \\
    3      & $-1.33$ \\
    4      & $-1.75$ \\
    8      & $-1.78$ \\
    51     & $-2.04$ \\
    39     & $-2.16$ \\
    6      & $-2.65$ \\
    33     & $-3.50$ \\
    5      & $-3.55$ \\
    2      & $-3.55$ \\
    1      & $-9.87$ \\
    \hline
  \end{tabular}
\end{table}

\begin{table}[htbp] 
\centering 
\caption{\ce{CO2} adsorption energies (eV) at the bridge (hollow) site of pristine single-metal reference MBenes computed in this work and compared with published theoretical values. The close agreement (mean absolute deviation below 0.15~eV) across five distinct MBene compositions validates the DFT protocol applied throughout the screening.} 
\label{tab:co2_ads_bridge} 
\begin{tabular}{lSS} 
\toprule {Pristine MBene} & {This work (eV)} & {Literature (eV)} \\ \midrule HfB & -2.71 & -2.61 \\ TiB & -3.10 & -3.27 \\ TaB & -1.48 & -1.44 \\ NbB & -1.68 & -1.75 \\ VB & -2.12 & -2.20 \\ \bottomrule \end{tabular} 
\end{table}

As shown in Figure~\ref{fig:ml}c, the global distribution of adsorption energies separates into two distinct populations: intact bent \ce{CO2^{\delta-}} configurations, which peak near $-2.0$~eV, and dissociated CO\,$+$\,O configurations, which peak near $-4.0$~eV. 
The crossover near $\Delta E_{\rm ads} \approx -2.5$~eV marks the activation--dissociation boundary and reflects the onset of antibonding $\sigma^*$ orbital population in the C--O bond under excessive electron transfer.

Figure~\ref{fig:ml}d further illustrates this boundary through the
relationship between the maximum C--O bond length and the adsorption energy, colour-coded by the O--C--O bond angle: intact configurations cluster near the equilibrium C--O bond length (dashed vertical line) with O--C--O angles that progressively bend from approximately $160^\circ$ toward $120^\circ$ as $|\Delta E_{\rm ads}|$ increases, while dissociated configurations extend to large C--O separations with angles below $60^\circ$.

\begin{figure}[tp!] 
\centering 
\includegraphics[width=0.85\linewidth]{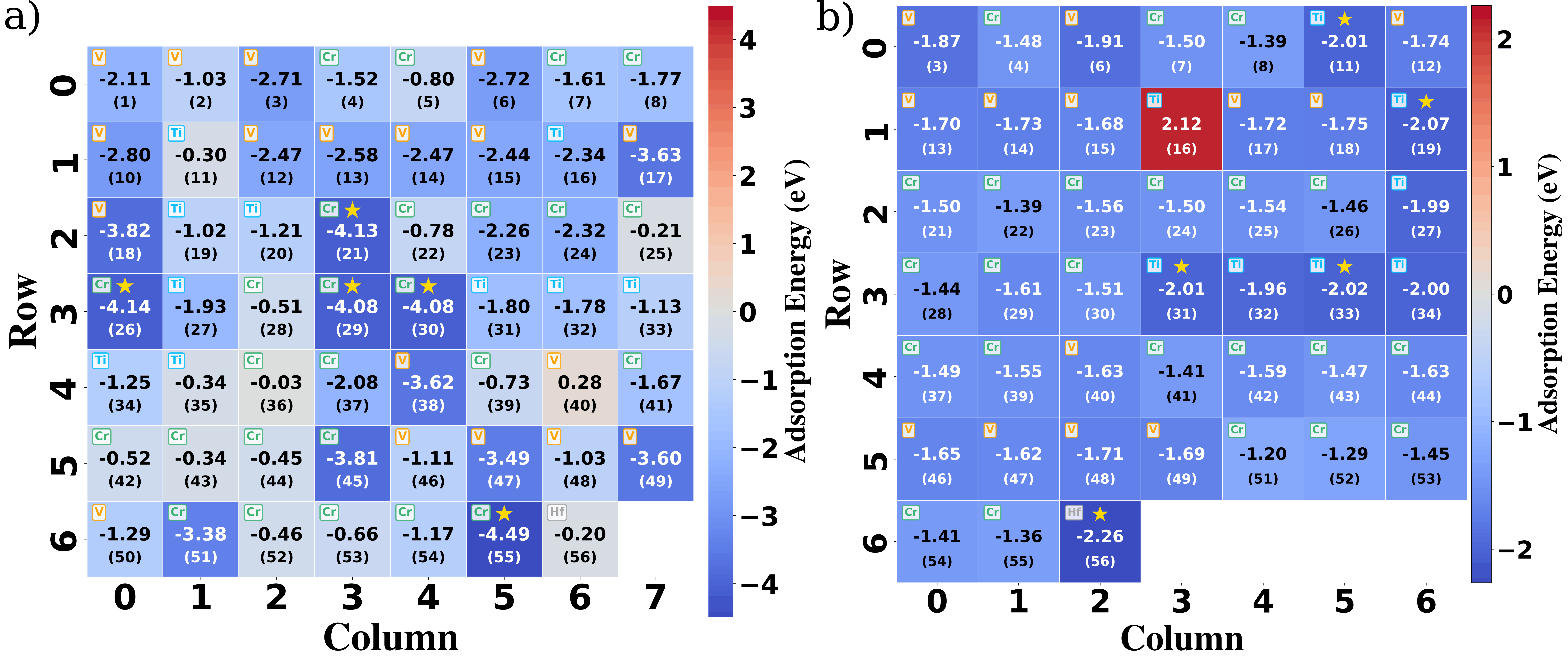} \caption{\textbf{DFT \ce{CO2} adsorption energy maps across the HE-MBene compositional space.} (a) On-top site \ce{CO2} adsorption energies ($\Delta E_{\rm ads}$, eV) for the primary subset of compositions. The PDOS-identified active adsorption element is labelled in each cell (V: vanadium; Cr: chromium; Ti: titanium; Hf: hafnium). Compositions where \ce{CO2} adsorbs dissociatively or with $\Delta E_{\rm ads} \geq 0$ are excluded from further analysis; stars ($\star$) mark global extrema. The three zero-overpotential (within the CHE thermodynamic framework) candidates (HES~IDs 22, 53, and 54) exhibit moderate Cr-centred binding energies of $-0.66$ to $-1.17$~eV, placing them within the Sabatier-optimal window for stepwise \ce{CO2}RR to CO. (b) {\textbf{\ce{CO2} adsorption at hollow (bridge) sites across the filtered HE-MBene compositional space.} Adsorption energies are reported for the complementary subset of compositions after excluding structures with positive formation or cohesive energies, completing coverage of the 56-member space. The three zero-overpotential candidates (HES~IDs 22, 53, and 54) exhibit moderate Cr-centred binding energies ($-0.66$ to $-1.17$~eV), consistent with the Sabatier-optimal window for stepwise \ce{CO2} reduction to CO.} 
}
\label{fig:co2_landscape} 
\end{figure}

A chemically coherent element-resolved pattern also emerges from inspection
of Figure~\ref{fig:co2_landscape}: Ti-centred sites consistently yield the
most negative adsorption energies (down to approximately $-4.1$~eV),
reflecting Ti's high Lewis acidity, while Cr- and V-centred configurations
display substantially weaker binding ($-0.30$ to $-2.0$~eV).
The three zero-overpotential compositions (HES~IDs 22, 53, and 54) exhibit
moderate Cr-centred adsorption energies of $-0.66$ to $-1.17$~eV, placing
them within the Sabatier-optimal binding window for
\ce{CO2}RR~\cite{Batchelor2019}: strong enough to activate \ce{CO2} and
stabilise \ce{COOH^*}, yet weak enough to allow facile desorption of
\ce{CO^*} and prevent catalyst poisoning.

Figure~\ref{fig:co2_landscape} maps the DFT \ce{CO2} adsorption energies throughout the compositional space, with the active adsorption element identified by PDOS labeled in each cell. Panel (a) covers the primary subset of 56 compositions at on-top sites; panel (b) completes the coverage of the remaining compositions. 
A clear element-specific hierarchy is visible: Ti-centred sites produce the most exothermic binding (global minimum near $-4.1$~eV), followed by Hf- and Zr-centred configurations, while Cr- and V-centred sites occupy a weaker-binding regime. The three zero-overpotential (within the CHE thermodynamic framework) candidates (HES~IDs 22, 53, 54) all show Cr-centred adsorption energies of $-0.66$ to $-1.17$~eV, placing them within the Sabatier-optimal window for selective stepwise reduction \ce{CO2} to CO. 
This moderate binding prevents dissociative chemisorption (characteristic of strongly exothermic Ti-centred configurations) and ineffective activation (characteristic of weakly binding V-rich surfaces), confirming that the Cr active site within the Hf/Zr donor matrix is the key chemical motif enabling zero-overpotential (within the CHE thermodynamic framework) performance.

\FloatBarrier 
\subsection{Free Energy Profiles and Zero-Overpotential Candidates}

\begin{figure}[tp!] 
\centering 
\includegraphics[width=0.85\linewidth]{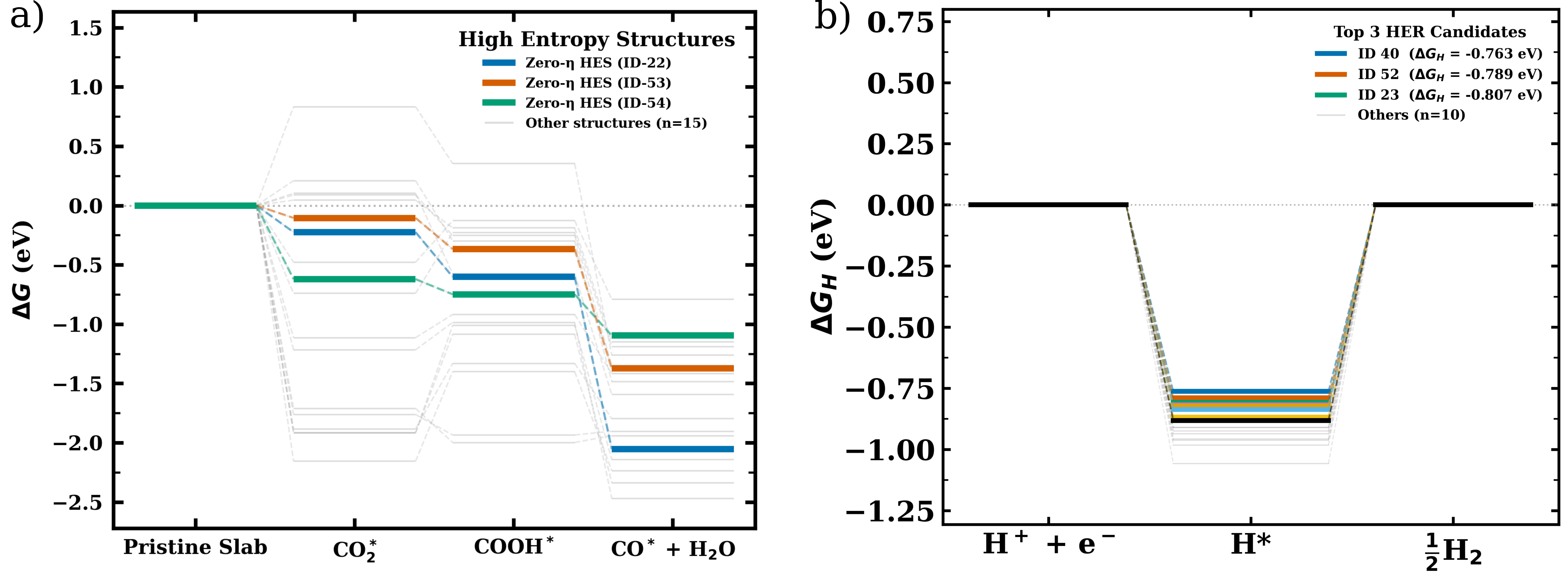} 
\caption{\textbf{CHE free energy profiles for \ce{CO2}RR and HER selectivity assessment at zero applied potential ($U = 0$~V vs.\ RHE).}
(a) Free energy profiles for the two-electron \ce{CO2}-to-CO pathway.
Coloured solid lines: the three zero-overpotential candidates (HES~ID-22, blue; HES~ID-53, orange; HES~ID-54, green), all exhibiting monotonically downhill profiles confirming $U_L = 0.00$~V vs.\ RHE. 
Grey lines: the remaining 15 compositions that passed the adsorption screening gate. The dashed horizontal reference at $\Delta G = 0$~eV denotes the thermodynamic threshold; uphill \ce{COOH^*} formation steps in grey profiles identify the characteristic rate-limiting step of aqueous \ce{CO2}RR. 
Reaction coordinates: pristine slab $\to$ \ce{CO2^*} $\to$ \ce{COOH^*} $\to$ \ce{CO^*} + \ce{H2O}. (b) Hydrogen adsorption free energy ($\Delta G_{H^*}$) profiles for the 18 final compositions. The top three candidates by HER activity (IDs 40, 52, and 23) are shown in colour alongside the remaining structures (grey). 
All compositions satisfy $\Delta G_{{\rm COOH}^*} < \Delta G_{H^*}$, confirming thermodynamic preference for \ce{CO2}RR over HER at the active site.} 
\label{fig:free} 
\end{figure}

Figure~\ref{fig:free} presents the CHE free energy profiles for the two-step \ce{CO2}-to-CO pathway at zero applied potential ($U = 0$~V vs.\ RHE) for all 18 compositions that survived the adsorption screening gate (panel a),together with the hydrogen adsorption profiles used to assess HER selectivity (panel b).
The reaction coordinate in panel (a) proceeds from left to right: pristine slab, \ce{CO2^*} adsorption, \ce{COOH^*} formation, and \ce{CO^*} + \ce{H2O} product release. The three zero-overpotential candidates (HES~IDs 22, 53, and 54) are highlighted in colour; the remaining 15 compositions are shown in grey. All three highlighted free energy profiles are monotonically downhill at every elementary step, confirming that no applied potential is thermodynamically required to drive \ce{CO2}-to-CO conversion. This outcome has not been reported for any previously characterised MBene or HEA catalyst for the two-electron \ce{CO2}-to-CO pathway. For context, the best-reported limiting potentials in the MBene literature are $-0.31$~V for \ce{Cr2B3} (toward \ce{CO2} to \ce{CH4})~\cite{peng2025}, $-0.46$~V for hydroxyl-terminated ScBOH h-MBene~\cite{di2025}, and $-0.27$ to $-0.32$~V for single-atom-decorated \ce{Mo2B2} MBene (toward \ce{CH3OH}
and \ce{CH4}, respectively)~\cite{bai2023}. These benchmarks target multi-electron products via distinct pathways, so the present zero-overpotential result for the two-electron \ce{CO2}-to-CO pathway represents a complementary and thermodynamically superior outcome on this specific reaction coordinate. The zero-overpotential performance of HES~IDs 22, 53, and 54 also surpasses the best reported HEA result, FeCoNiCuMo ($U_L = -0.29$ to $-0.51$~V, toward multi-electron products)~\cite{chen2022}, establishing HE-MBenes as the most thermodynamically efficient catalysts computationally identified for selective \ce{CO2}-to-CO conversion to date.
The selectivity over the competing HER is confirmed by panel (b) of Figure~\ref{fig:free}: the hydrogen adsorption free energy $\Delta G_{H^*}$ lies below $-0.76$~eV for all compositions that passed the final CHE gate, indicating that \ce{H^*} binds strongly but the \ce{COOH^*} formation step remains more favourable than the Volmer step across all 18 candidates. This satisfies the criterion $\Delta G_{{\rm COOH}^*} < \Delta G_{H^*}$ and rules out HER as the dominant pathway at the active site under zero-bias conditions.
Among the three candidates, HES~ID-54 achieves the most balanced thermodynamic distribution: $\Delta G = -1.17$~eV (\ce{CO2^*} adsorption), $-0.74$~eV (\ce{COOH^*} formation), and $-1.09$~eV (\ce{CO^*} + \ce{H2O} release). This relative uniformity of exergonicity minimises the thermodynamic driving-force gradient and the risk of intermediate trapping, which can lead to catalyst deactivation through \ce{CO^*} poisoning or accumulation of \ce{COOH^*}. HES~ID-53 presents a similarly balanced profile ($\Delta G_{\rm CO^*+H_2O} = -1.37$~eV), making it the second most promising candidate from a turnover-robustness perspective.
HES~ID-22, while showing a strongly spontaneous terminal step ($\Delta G = -2.05$~eV for \ce{CO^*} + \ce{H2O}), carries a risk of elevated CO desorption barriers under cycling conditions due to deep stabilisation of \ce{CO^*}; modest applied potential or temperature may be needed to achieve sustained turnover in practice.
The 15 grey trajectories show an uphill \ce{COOH^*} formation step as the characteristic rate-limiting barrier under aqueous \ce{CO2}RR conditions~\cite{norskov2004,di2025}. 

Table~\ref{tab:top_candidates} summarises the key electronic and thermodynamic descriptors for the three zero-overpotential candidates. Their moderate Cr-centred \ce{CO2} adsorption energies ($-0.66$ to $-1.17$~eV) are consistent with the Sabatier optimum~\cite{Batchelor2019, pedersen2020}. The zero limiting potential ($U_L = 0.00$~V) arises because high-entropy compositional disorder simultaneously optimises $\Delta G_{\rm COOH^*}$ and $\Delta G_{\rm CO^*+H_2O}$, a feat thermodynamically impossible on monometallic surfaces where LSRs couple these quantities through the universal CO-binding energy descriptor~\cite{Nellaiappan2020,chen2022}. 
The near-optimal $d$-band centres ($\varepsilon_d \approx -2.59$ to $-2.69$~eV) are direct electronic consequences of the Hf/Zr electron-donor framework: these elements elevate the local electron density at Cr active sites, shifting $\varepsilon_d$ toward the Fermi level and lowering the charge-transfer barrier to the \ce{CO2} $\pi^*$ orbital.

\begin{table}[!htbp] \centering \caption{\textbf{Key electronic and thermodynamic descriptors for the three zero-overpotential HE-MBene candidates.} All adsorption and free energy values in eV; $d$-band centre ($\varepsilon_d$) relative to $E_F$ (eV); limiting potential $U_L$ in V vs.\ RHE. 
Active sites are identified by element-resolved $d$-PDOS analysis as described in the text. All three candidates share Cr as the active site, with Hf/Zr electron-donor neighbours providing local charge enrichment.} \label{tab:top_candidates} 
\small 
\begin{tabular}{lllccccc} 
\hline HES & Composition & Active & $E_{{\rm CO}_2^*}$ & $\Delta G_{{\rm COOH}^*}$ & $\Delta G_{{\rm CO}^*+{\rm H_2O}}$ & $\varepsilon_d$ & $U_L$ \\ ID & & Site & (eV) & (eV) & (eV) & (eV) & (V) \\ \hline 22 & CrNbZrMoTiB$_5$ & Cr & $-0.78$ & $-0.60$ & $-2.05$ & $-2.64$ & 0.00 \\ 53 & MoZrHfNbCrB$_5$ & Cr & $-0.66$ & $-0.36$ & $-1.37$ & $-2.69$ & 0.00 \\ 54 & MoZrHfTaCrB$_5$ & Cr & $-1.17$ & $-0.74$ & $-1.09$ & $-2.59$ & 0.00 \\ 
\hline 
\end{tabular} 
\end{table} \FloatBarrier 

\subsection{Atomic-Scale Verification of the Rate-Determining Step} 
\begin{figure}[tp!] 
\centering 
\includegraphics[width=\linewidth]{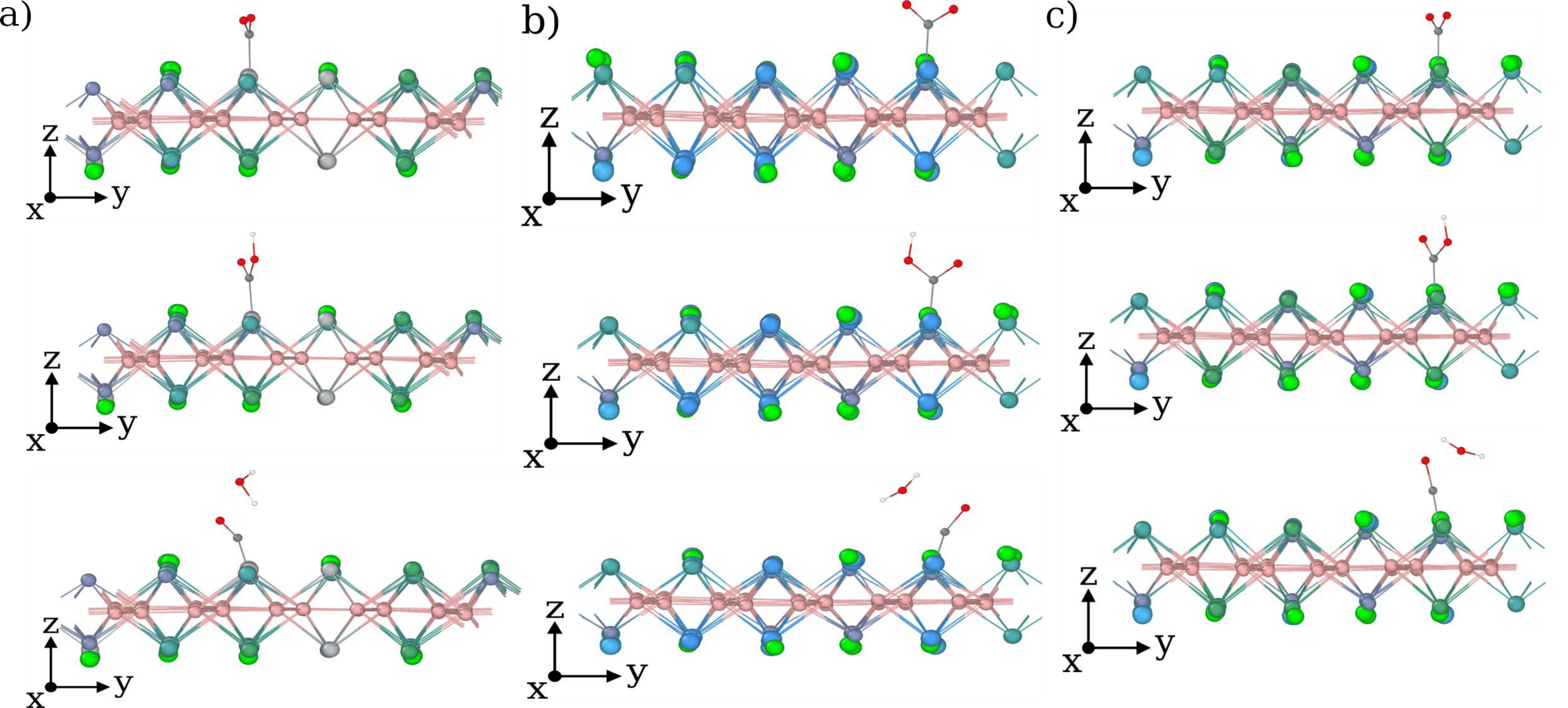} 
\caption{\textbf{Atomic snapshots of the rate-determining \ce{COOH^*} intermediate on the three zero-overpotential HE-MBene catalysts.} Columns (a)--(c): HES~ID-22, HES~ID-53, and HES~ID-54, respectively. Rows 1--3: three sequential side-view orientations (translated along $z$) revealing the monodentate adsorption geometry and the local elemental environment surrounding the active site. Red sphere: O; small grey sphere: H; pink spheres: B; large coloured spheres: metal atoms (colour-coded per composition legend). The \ce{COOH^*} fragment binds through the carboxyl C to the Cr active site ($d_{\rm C-Cr} = 2.01$--$2.14$~\AA{}); adjacent Hf/Zr donor atoms are visible as nearest-neighbour metal atoms in the top layer, providing continuous charge enrichment to the active site. Coordinate axes are shown in each panel.} 
\label{fig:rds} 
\end{figure}

Figure~\ref{fig:rds} presents atomic snapshots of the DFT-relaxed \ce{COOH^*} intermediate geometries on HES~IDs 22, 53, and 54 (columns a, b, and c, respectively) at three sequential $z$-translations (rows 1--3), revealing both the molecular adsorption geometry and the local elemental environment of the active site.
The \ce{COOH^*} fragment adopts an upright monodentate configuration in all nine panels, with carboxyl carbon coordinated to a Cr atom and the hydroxyl oxygen directed toward the vacuum. The C-Cr bond lengths extracted from converged DFT geometries fall in the range 2.01--2.14~\AA{}, consistent with strong chemisorption without the bond strain that would indicate incipient dissociation. 

These values agree well with the C-Cr distances reported for \ce{COOH^*} on a single-element \ce{Cr2B3} (approximately 2.05~\AA{}~\cite{peng2025}), confirming that the active Cr site retains its fundamental chemical character within the high-entropy matrix. Critically, the MBene lattice shows no significant structural distortion in the immediate vicinity of the Cr adsorption site across all three compositions, confirming that the zero-overpotential performance does not rely on local lattice contraction, an important criterion for catalytic robustness under sustained electrochemical cycling. 

The color coding of metal atoms makes the spatial relationship between the Hf/Zr donor elements and the Cr acceptor directly visible: large Hf and Zr atoms occupy nearest-neighbour metal sites in the upper layer, placing them within the Bader charge redistribution radius of the bound \ce{COOH^*} fragment. This spatial arrangement directly validates the mechanistic picture derived from the Bader charge and PDOS analyses: Hf/Zr-mediated electron donation continuously enriches the local electron density at the Cr active site, stabilizing the \ce{COOH^*} intermediate through enhanced back-donation and producing the spontaneously downhill free energy profiles observed in Figure~\ref{fig:free}. 
\FloatBarrier 

\subsection{Thermal Stability: AIMD Simulations} 
\begin{figure}[H] 
\centering 
\includegraphics[width=0.85\linewidth]{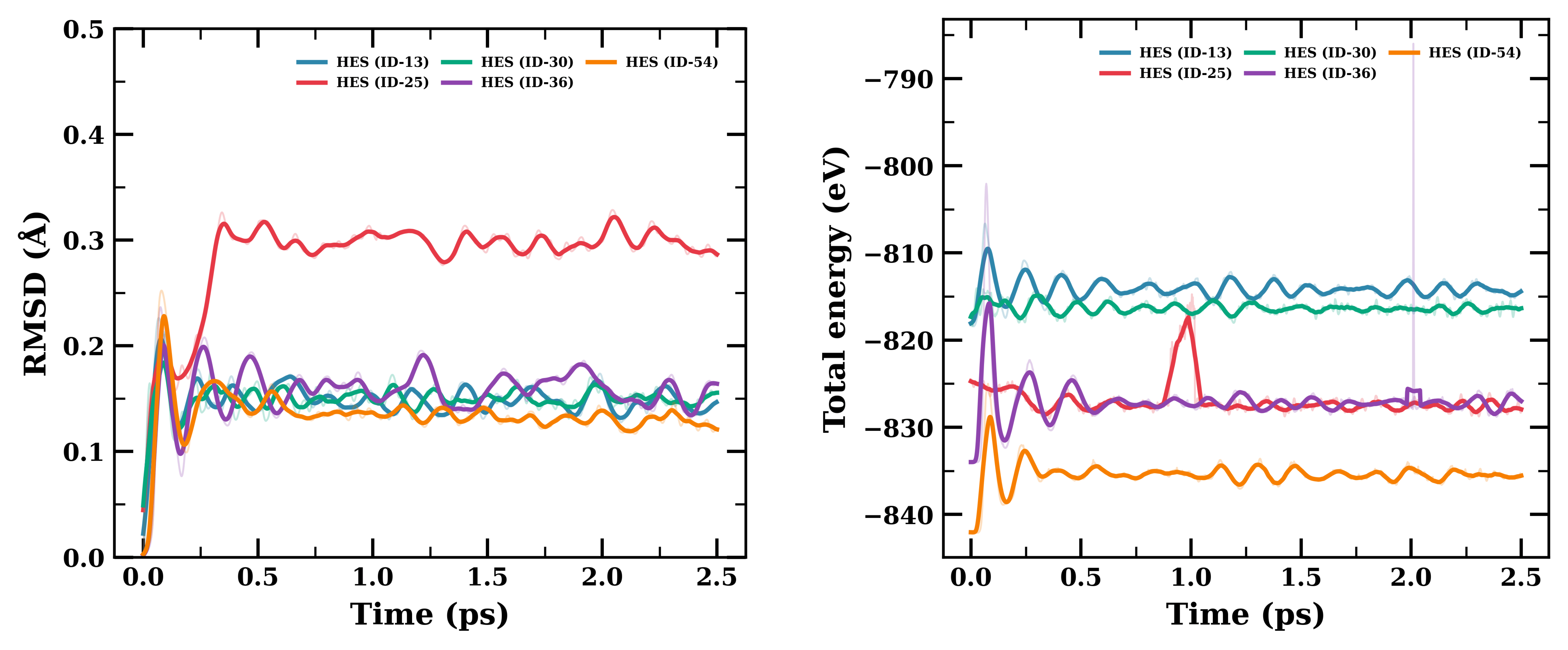} \caption{\textbf{AIMD thermal stability analysis at 500~K for five representative HE-MBene compositions.} 
(a) Root-mean-square displacement (RMSD, \AA{}) as a function of simulation time (ps) under NVT conditions. Plateau RMSD values of 0.12--0.20~\AA{} for HES~IDs 13, 30, 36, and 54 are consistent with equilibrium lattice vibrations; the absence of a monotonically rising RMSD rules out amorphisation or surface reconstruction. HES~ID-36 is included as a near-stability-boundary benchmark despite its positive formation energy (see text). (b) Total energy (eV) evolution over the trajectory; convergence within approximately 0.5~ps for all compositions confirms thermal equilibration. 
HES~ID-54 achieves the most negative stable energy, consistent with its high cohesive stability.} 
\label{fig:aimd} 
\end{figure}

Figure~\ref{fig:aimd} presents the evolution of RMSD and total energy on 2.5~ps NVT AIMD trajectories at 500~K for five representative compositions (HES~IDs 13, 25, 30, 36 and 54). 

The trajectories reveal rapid initial structural relaxation within the first 0.3~ps for four of the five compositions (HES~IDs 13, 30, 36, and 54), after which the RMSD plateaus between 0.12 and 0.20~\AA{}. 
These values are characteristic of equilibrium thermal vibrations on a stable 2D lattice and are comparable to values reported for thermally robust MXene and MBene surfaces at similar temperatures~\cite{lu2023}. HES~ID-36, though excluded from the catalytic screening funnel on thermodynamic grounds ($\Delta E_f = +0.13$~eV/atom), was retained for AIMD benchmarking to assess whether kinetically accessible metastable compositions remain structurally robust at elevated temperature; its plateau RMSD of approximately 0.16~\AA{} confirms that even this near-stability-boundary composition sustains lattice integrity at 500~K. 
HES~ID-25 reaches a slightly elevated steady-state RMSD of approximately 0.29~\AA{}, indicating greater thermal disorder; the clear plateau character confirms that this represents sustained dynamic equilibration rather than irreversible structural decomposition. 
None of the five compositions exhibits a monotonically rising RMSD, which is the diagnostic signature of amorphisation or surface reconstruction, confirming mechanical robustness at temperatures well above typical aqueous \ce{CO2}RR operating conditions (below 353~K). 

Total energies converge to stable averages within approximately 0.5~ps for all five compositions; HES~ID-54 stabilizes at the most negative value (approximately $-838$~eV), consistent with its strong cohesive energy, confirming that the most catalytically active composition is also among the most thermally stable. 

\begin{figure}[tp!] 
\centering 
\includegraphics[width=0.72\linewidth]{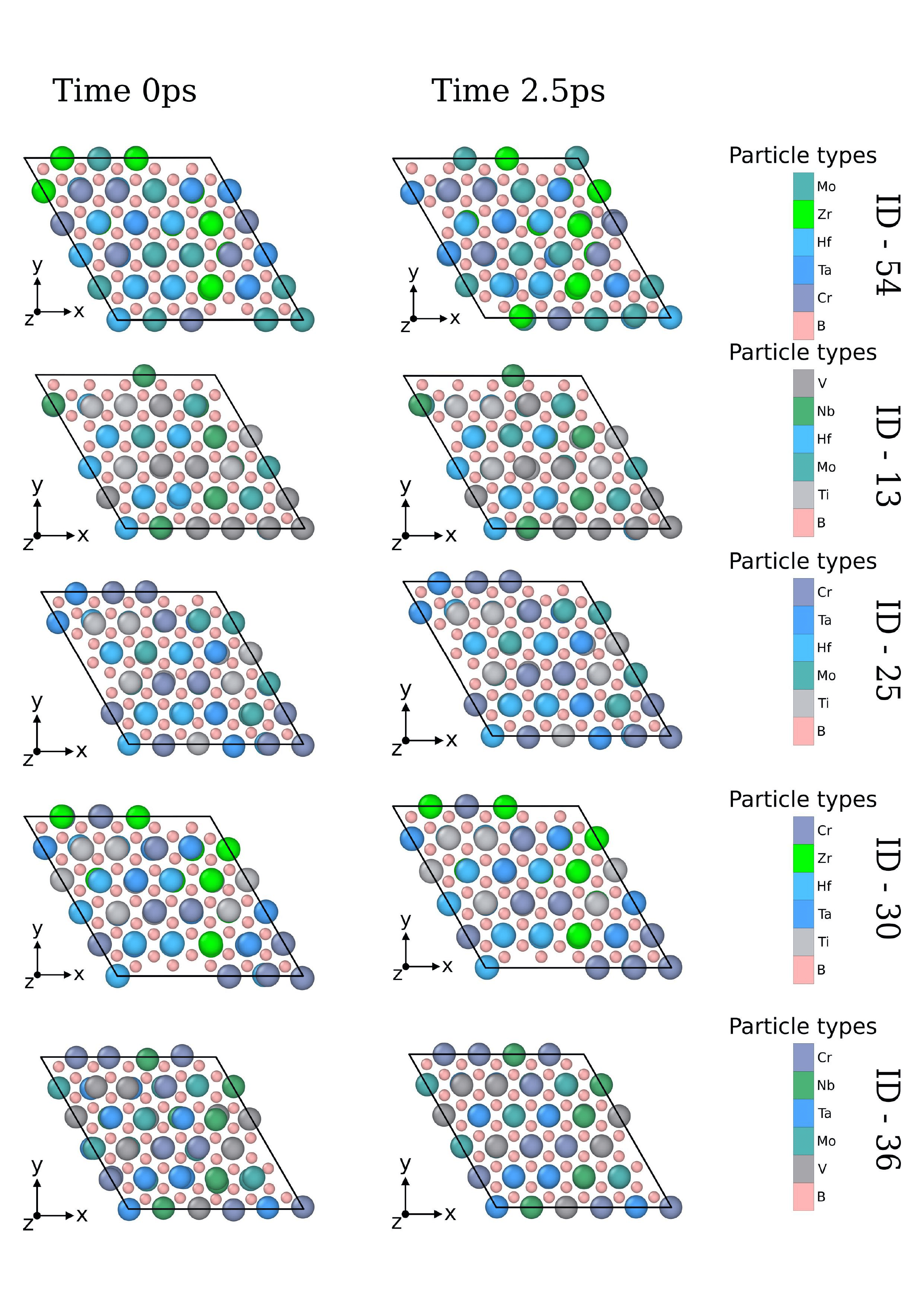} 
\caption{\textbf{Atomic snapshots at the start and end of AIMD trajectories at 500~K confirming lattice preservation.} Top-view supercell snapshots at $t = 0$~ps (left column) and $t = 2.5$~ps (right column) for HES~IDs 54, 13, 25, 30, and 36 (rows top to bottom). Atom colour coding per individual legends; boron atoms are pink. The hexagonal M$_1$B$_1$ lattice topology is preserved throughout in all five compositions, with no atomic desorption, phase separation, or cation-anion redistribution, confirming the absence of irreversible structural changes at 500~K.} 
\label{fig:snaps} 
\end{figure}

Figure~\ref{fig:snaps} provides complementary atomic-scale evidence through top-view supercell snapshots at $t = 0$ and $t = 2.5$~ps for all five compositions. The characteristic hexagonal MBene lattice topology is fully preserved throughout the trajectory in all five cases, with no evidence of atomic desorption, vacancy formation, phase separation, or significant cation-anion site redistribution. 
This confirms that the RMSD plateaus correspond exclusively to equilibrium thermal vibrations and not to irreversible structural changes, validating HES~IDs 13, 30, and 54 as thermally robust candidates for experimental synthesis and electrochemical characterization. 

\FloatBarrier 

\subsection{Structure-Property Relationships and Compositional Design Principles} The convergence of evidence from property maps, PDOS and work function data, the 1,375-configuration adsorption landscape, CHE free energy profiles, RDS geometries, and AIMD trajectories establishes a coherent and self-consistent set of structure-property relationships for HE-MBene \ce{CO2}RR catalysts. Three overarching compositional design principles emerge, each independently supported by multiple lines of computational evidence. 

\textbf{Principle I: Electron-donor enrichment governs thermodynamic stability and \ce{CO2} activation.} Compositions enriched in group-IV elements (Hf, Zr, Ti) consistently exhibit the most negative formation energies, the lowest work functions, and the most exothermic \ce{CO2} adsorption energies. 
Element-resolved Bader charge analysis identifies Hf and Zr as the primary electron donors in the mixed metal sublattice. 
Their charge donation is most catalytically effective when directed toward the active Cr site, whose $d$-PDOS in the $-2$ to $+1$~eV window relative to $E_F$ has the largest unoccupied weight among the eight candidate elements for all three zero-overpotential compositions (Figure~\ref{fig:pdos_wf}). 
Thus, a necessary condition for zero-overpotential performance emerges: the composition must include at least one Hf or Zr donor element and must carry Cr as the preferred adsorption center. 

\textbf{Principle II: High-entropy configurational disorder decouples intermediate scaling relationships.} 
The zero-overpotential performance of HES~IDs 22, 53 and 54 cannot be rationalized within the universal LSR framework governing monometallic surfaces, wherein $\Delta G_{\rm COOH^*}$ and $\Delta G_{\rm CO^*+H_2O}$ are thermodynamically coupled through the CO binding energy descriptor. 
The diverse local chemical environments generated by the equiatomic five-element mixing create a heterogeneous free energy landscape in which both quantities are independently optimized. This constitutes direct computational evidence at the 2D MBene level of the LSR-breaking cocktail effect first proposed for bulk HEAs~\cite{pedersen2020,chen2022}. 
The physical origin lies in the absence of a single, well-defined surface electronic descriptor when five elements are mixed equiatomically: adsorbate interactions at individual sites are governed by their specific local neighborhood rather than by a global scaling law. 

\textbf{Principle III: Balanced compositions outperform extremes.} The two most active and thermally stable candidates (HES~ID 53 and 54) share a combination of Hf/Zr/Mo/Cr that simultaneously provides electron donation (Hf, Zr), intermediate $d$-band positioning (Mo), and Lewis-acid reactivity at the active site (Cr). 
Compositions dominated exclusively by strong donors (Hf/Zr/Ti-rich without Cr) overbind \ce{CO2} and produce C-O dissociation, preventing controlled stepwise \ce{COOH^*} formation. 
In contrast, V/Cr-dominated compositions without sufficient group-IV donors show insufficient \ce{CO2} activation. The optimal adsorption window of $-0.66$ to $-1.17$~eV for candidates with zero-overpotential is in excellent agreement with the Sabatier optimum for the two-electron \ce{CO2}RR to CO~\cite{Batchelor2019}, providing a quantitative target for future exploration. 
Together, these three principles define a concise and actionable compositional design recipe: combine Hf or Zr electron donors with a Cr active site within a Mo/Nb/Ta intermediate-field matrix to achieve balanced \ce{CO2} activation, optimal stabilization of \ce{COOH^*}, and facile \ce{CO^*} desorption.

\section{Conclusion}
We have presented the first systematic, multi-fidelity computational screening of high-entropy MBenes as electrochemical \ce{CO2} reduction catalysts. Our study covered all 56 equiatomic quinary compositions accessible from the \{Ti, V, Cr, Mo, Nb, Ta, Zr, Hf\} elemental pool. 
To efficiently navigate this vast compositional space, we employed a five-stage sequential screening funnel comprising structural relaxation and mechanical stability assessment, thermodynamic formation energy filtering, phonon dynamical stability screening, density functional theory (DFT) \ce{CO2} adsorption screening guided by PDOS-identified active sites, and computational hydrogen electrode (CHE) rate-determining step analysis. 
The integration of a MACE machine-learning interatomic potential ($R^2 = 0.9991$; RMSE $= 3.49$~meV/atom) enabled computationally efficient adsorption evaluation across 1,375 unique surface environments, ultimately reducing the initial space from 56 down to 18 viable candidates.
These 18 candidate structures demonstrate robust stability and favorable thermodynamics for catalytic applications. Ab initio molecular dynamics (AIMD) simulations conducted at 500~K confirm thermal stability, showing minimal root-mean-square deviations (RMSD 0.12--0.20~\AA{}) without structural degradation. 
Crucially, selectivity for the \ce{CO2} reduction reaction over the competing hydrogen evolution reaction (HER) was successfully verified, with all 18 candidates satisfying the thermodynamic condition $\Delta G_{\ce{COOH}^*} < \Delta G_{\ce{H}^*}$.
Among the viable candidates, three compositions emerged with exceptional zero-overpotential \ce{CO2}-to-CO performance:\ce{CrNbZrMoTiB5}, \ce{MoZrHfNbCrB5}, and \ce{MoZrHfTaCrB5}. All three exhibit spontaneously downhill free energy profiles at $U = 0$~V vs.\ RHE. This remarkable performance surpasses all reported conventional MBene and high-entropy alloy benchmarks for the two-electron \ce{CO2}-to-CO pathway.
Our atomistically complete and internally consistent mechanistic analysis reveals the underlying principles driving this catalytic activity. Bader charge analysis, element-resolved $d$-PDOS, and electrostatic potential profiles demonstrate that Hf and Zr elevate the local electron density at Cr-centered active sites. 
This is further corroborated by relaxed \ce{COOH^*} geometries exhibiting C–-Cr bond lengths of 2.01--2.14~\AA{}. Consequently, the $d$-band center shifts toward the Fermi level ($E_F$), stabilizing the rate-determining \ce{COOH^*} intermediate via enhanced back-donation. 
The high-entropy configurational disorder plays a vital role here, decoupling linear scaling relationships and enabling independent optimization of all elementary step free energies. From these insights, we derive a concise compositional design rule: combine Hf or Zr electron donors with a Cr active site within a Mo/Nb/Ta intermediate-field matrix to achieve optimally balanced \ce{CO2} activation and product desorption.

While these computational findings are highly promising, several avenues remain for exploration. Future directions include experimental synthesis via quinary MAB-phase etching alongside structural characterization, and electrochemical validation in realistic environments. 
Additionally, extending this methodology to higher-order HE-MBene systems and multi-product \ce{CO2}RR pathways presents an exciting frontier. Ultimately, our integrated workflow---spanning MCSQS, DFT, PDOS analysis, MACE-MLIP screening, CHE modeling, and AIMD validation---provides a scalable paradigm for accelerated discovery of high-entropy two-dimensional materials.

\section{Funding Declaration}
This work did not receive any funding. 
\begin{acknowledgement}
We acknowledge the \emph{Param Ananta} supercomputing facility at IIT Gandhinagar, supported by the National Supercomputing Mission for providing computing resources for all the simulations reported in this work. Helpful discussions with Dr. Kabeer Jasuja, IIT Gandhinagar are acknowledged. 
\end{acknowledgement}

\bibliography{references}

\end{document}